\documentclass [ 10 pt ] {article}
\usepackage{tikz, pgf, pgfplots}
\usetikzlibrary
	{
		3d,
		arrows,									
		arrows.meta,
		automata,								
		backgrounds, 
		bending, 
		calc, 
		chains, 
		datavisualization.formats.functions,		
		decorations.pathmorphing,
		decorations.pathreplacing,
		decorations.text,
		fadings,
		fit, 
		graphs,
		graphs.standard,
		matrix,
		positioning,								
		quantikz2,
		quotes,
		shadings,
		shadows.blur,
		shapes, 
		trees
	}
\pgfplotsset{compat = newest}
\usepgflibrary
	{
		shapes.misc
	}
\usepgfplotslibrary
	{
		dateplot
	}
\usepackage{tikzpeople}
\usepackage{fontawesome5}

\usepackage{amsthm}
\usepackage{amssymb}
\usepackage[bb = boondox]{mathalfa}
\usepackage{dsfont}
\usepackage{newtxmath}
\usepackage{mathtools}
\usepackage{multicol}
\usepackage{braket}
\usepackage{physics}

\usepackage[ compat = newest ]{yquant}

\usepackage{xcolor}
\usepackage{varwidth}
\usepackage[ skins, breakable, listings, theorems ] {tcolorbox}
\usepackage{graphicx}
\usepackage{xurl}                                                                                  %
\usepackage{hyperref}
\hypersetup
	{
		colorlinks,
		linkcolor = {WordRed!75!black},
		citecolor = {WordBlueDarker25},
		urlcolor = {WordAquaDarker50}
	}
\usepackage{colortbl}
\usepackage{float}
\usepackage{nicematrix}
\usepackage{cellspace}
\usepackage{makecell}
\usepackage{multirow}
\usepackage[ linesnumbered, tworuled, vlined ] {algorithm2e}
\usepackage{appendix}
\usepackage{enumitem}
\usepackage{xintexpr}
\usepackage{spreadtab}
\usepackage{framed}
\usepackage{lipsum}
\usepackage{orcidlink}                                                                           %
\newcommand{\orcidicon}[1]{\href{https://orcid.org/#1}{\includegraphics[height=\fontcharht\font`\B]{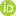}}}

\newtheorem{definition}{Definition}[section]

\newtheorem{example}{Example}[section]

\newcounter{mathseed}
\pgfmathsetseed{\arabic{mathseed}}
\pgfdeclarelayer{background}
\pgfsetlayers{background, main}
\pgfdeclaredecoration{irregular fractal line}{init}
{
	\state{init}[width=\pgfdecoratedinputsegmentremainingdistance]
	{
		\pgfpathlineto{\pgfpoint{random*\pgfdecoratedinputsegmentremainingdistance}{(random*\pgfdecorationsegmentamplitude-0.02)*\pgfdecoratedinputsegmentremainingdistance}}
		\pgfpathlineto{\pgfpoint{\pgfdecoratedinputsegmentremainingdistance}{0pt}}
	}
}
\def\tornpaper#1{%
	\ifthenelse{\isodd{\value{mathseed}}}
	{%
		\tikz
		{
			\node[inner sep = 1em] (A) {#1};		
			\begin{pgfonlayer}{background}			
				\fill[paper]						
				\pgfextra{\pgfmathsetseed{\arabic{mathseed}}\addtocounter{mathseed}{1}}%
				{decorate[irregular cloudy border]{decorate{decorate{decorate{decorate[ragged border]{
										(A.north west) -- (A.north east)
				}}}}}}
				-- (A.south east)
				\pgfextra{\pgfmathsetseed{\arabic{mathseed}}}%
				{decorate[irregular spiky border]{decorate{decorate{decorate{decorate[ragged border]{
										-- (A.south west)
				}}}}}}
				-- (A.north west);
			\end{pgfonlayer}
		}
	}
	{%
		\tikz{
			\node[inner sep=1em] (A) {#1};  
			\begin{pgfonlayer}{background}  
				\fill[paper] 
				\pgfextra{\pgfmathsetseed{\arabic{mathseed}}\addtocounter{mathseed}{1}}%
				{decorate[irregular spiky border]{decorate{decorate{decorate{decorate[ragged border]{
										(A.north east) -- (A.north west)
				}}}}}}
				-- (A.south west)
				\pgfextra{\pgfmathsetseed{\arabic{mathseed}}}%
				{decorate[irregular cloudy border]{decorate{decorate{decorate{decorate[ragged border]{
										-- (A.south east)
				}}}}}}
				-- (A.north east);
		\end{pgfonlayer}}
	}
}
\definecolor{MyLightRed}{RGB}{244, 213, 245}
\definecolor{WordRed}{RGB}{255, 0, 102}
\definecolor{RedDarkLightest}{HTML}{ff0088}
\definecolor{RedDarkLight}{HTML}{ea005f}
\definecolor{RedPurple}{HTML}{aa007f}
\definecolor{Purple}{HTML}{911146}
\definecolor{PurpleDark}{RGB}{102, 0, 102}
\definecolor{WordLightGreen}{RGB}{140, 214, 192}
\definecolor{WordGreen}{RGB}{0, 176, 80}
\definecolor{GreenLightest}{HTML}{00ffa0}
\definecolor{GreenLighter1}{HTML}{00b383}
\definecolor{GreenLighter2}{HTML}{00aa7f}
\definecolor{GreenDark}{HTML}{225522}
\definecolor{GreenTeal}{HTML}{008080}
\definecolor{WordIceBlue}{RGB}{223, 227, 229}
\definecolor{MyVeryLightBlue}{RGB}{211, 245, 247}
\definecolor{WordBlueVeryLight}{RGB}{0, 176, 240}
\definecolor{WordBlueLight}{RGB}{0, 112, 192}
\definecolor{WordBlueDark}{RGB}{46, 116, 181}
\definecolor{WordBlueDarker}{RGB}{31, 78, 121}
\definecolor{WordBlueDarker25}{RGB}{54, 96, 146}
\definecolor{WordBlueDarker50}{RGB}{36, 64, 98}
\definecolor{WordBlueDarkest}{RGB}{0, 32, 96}
\definecolor{WordBlue}{RGB}{19, 65, 99}
\definecolor{MyBlue}{RGB}{0, 64, 128}
\definecolor{MyDarkBlue}{RGB}{0, 51, 102}
\definecolor{BlueVeryDark}{HTML}{222255}
\definecolor{MagentaVeryLight}{RGB}{178, 162, 201}
\definecolor{MagentaLighter}{RGB}{161, 106, 221}
\definecolor{MagentaLight}{RGB}{128, 100, 162}
\definecolor{MagentaDark}{RGB}{106, 65, 152}
\definecolor{MagentaVeryDark}{RGB}{97, 75, 128}
\definecolor{WordAquaLighter80}{RGB}{218, 238, 243}
\definecolor{WordAquaLighter60}{RGB}{183, 222, 232}
\definecolor{WordAquaLighter40}{RGB}{146, 205, 220}
\definecolor{WordAquaDarker25}{RGB}{49, 134, 155}
\definecolor{WordAquaDarker50}{RGB}{33, 89, 103}
\definecolor{WordVeryLightTeal}{RGB}{223, 236, 235}
\definecolor{WordLightTeal}{RGB}{160, 199, 197}
\definecolor{WordDarkTealLighter80}{RGB}{207, 223, 234}
\definecolor{WordDarkTeal}{RGB}{72, 123, 119}
\definecolor{WordDarkerTeal}{RGB}{48, 82, 80}
\definecolor{WordTurquoiseLighter80}{RGB}{209, 238, 249}
\definecolor{Brown}{HTML}{666633}

\usepackage[a4paper, left = 2.5cm, right = 2.5 cm, top = 2.5cm, bottom = 3.0 cm]{geometry}
\title
	{
		A Novel Scalable Quantum Protocol for the Dining Cryptographers Problem
	}

\author
	{
		Peristera Karananou\orcidicon{0009-0007-5315-4600}
		and
		Theodore Andronikos\orcidicon{0000-0002-3741-1271}
		\\
		Department of Informatics, Ionian University, \\
		7 Tsirigoti Square, 49100 Corfu, Greece; \\
		\{p18kara4, andronikos\}@ionio.gr
	}
\begin{document}

\maketitle

\begin{abstract}
	This paper presents an innovative entanglement-based protocol to address the Dining Cryptographers problem, utilizing maximally entangled $\ket{ GHZ_{ n } }$ tuples as its core. This protocol aims to provide scalability in terms of both the number of cryptographers $n$ and the amount of anonymous information conveyed, represented by the number of qubits $m$ within each quantum register. The protocol supports an arbitrary number of cryptographers $n$, enabling scalability in both participant count and the volume of anonymous information transmitted. While the original Dining Cryptographers problem focused on a single bit of information – whether a cryptographer paid for dinner – the proposed protocol allows $m$, the number of qubits in each register, to be any arbitrarily large positive integer. This flexibility permits the conveyance of various information, such as the cost of the dinner or the timing of the arrangement. Another noteworthy aspect of the introduced protocol is its versatility in accommodating both localized and distributed versions of the Dining Cryptographers problem. The localized scenario involves all cryptographers gathering physically at the same location, such as a restaurant, simultaneously. In contrast, the distributed scenario accommodates cryptographers situated in different places, engaging in a virtual dinner at the same time. Finally, in terms of implementation, the protocol ensures uniformity by requiring all cryptographers to utilize identical private quantum circuits. This design establishes a completely modular quantum system where all modules are identical. Furthermore, each private quantum circuit exclusively employs the widely used Hadamard and CNOT quantum gates, facilitating straightforward implementation on contemporary quantum computers.
\end{abstract}
\section{Introduction} \label{sec: Introduction}

In today’s rapidly evolving digital era, technology seamlessly integrates into every aspect of our lives, a fact that makes cybersecurity more crucial than ever. As we navigate the complexity of digital boundaries, we find ourselves immersed in a world where unforeseen threats coexist with the free exchange of knowledge. The concept of privacy has evolved, encompassing critical aspects of individual freedom and security in an era characterized by rapid technological discoveries and digitally interconnected realms. In such a time, privacy refers to an individual's ability to manage personal information, choosing whether and how it is collected, used, and disclosed. The spectrum of issues where privacy is considered necessary has expanded significantly, ranging from private communications and financial transactions to private health information.

Privacy issues have become more intricate as digital platforms and smart technologies have become more ingrained in our daily lives. Social media, internet services, and data-driven technologies have brought unprecedented convenience, triggering urgent concerns about individual privacy. The clash between the need for innovation and the demand for personal data protection has emerged as a central topic of discussion in recent years.

Cybersecurity is entrusted with safeguarding our data, privacy, and networked culture. It is responsible for protecting digital systems, networks, and devices from various threats in cyberspace. These risks, ranging from ransomware attacks and advanced espionage to computer viruses and data breaches, not only pose economic risks but also jeopardize the foundation of trust on the Internet. With the prevailing evolution, even human life is at stake. The development of the Internet and the widespread use of smart gadgets have transformed the way we work, communicate, shop, and spend our leisure time. However, the digital revolution has increased the attack surface that cybercriminals can exploit to their advantage. Consequently, cybersecurity is a constantly evolving field to stay ahead of these risks. Alongside the evolution of software and hardware, the security of these systems must also progress, fortifying each system against new threats. This is achieved by combining technological advancements, regulations, and human skills. As we traverse the realm of cybersecurity, it becomes evident that safeguarding our digital future involves more than simple data protection. Understanding and implementing robust cybersecurity measures are key to building a safer, more resilient digital society.

The cryptographic protocol known as the Dining Cryptographers Problem explores the idea of anonymous communication in a social context. David Chaum first introduced it in 1988 \cite{Chaum1988} as a thought experiment to demonstrate the possibilities of private and secure communication among participants. Emphasis is placed on preserving the privacy and anonymity of each participant to achieve the goal of exchanging messages. To prevent the revelation of individual contributions, the protocol uses cryptographic techniques to ensure that the information exchanged between cryptographers reveals only the pre-agreed result ($0$ / $1$). The scenario is based on real-life situations where people desire to share information with others while maintaining their privacy and the confidentiality of their messages.


Boykin in 2002 proposed a quantum protocol, where participants distribute pairs of entangled qubits known as EPR pairs, which are subsequently utilized to generate cryptographic keys \cite{Boykin2002}. An EPR pair consists of two qubits entangled in a maximally entangled state, serving as a valuable resource for quantum communication and quantum computation, such as quantum teleportation. Boykin's system enables anonymous transmission of classical information based on quantum teleportation.
Subsequently, Christandl and Wehner introduced a new protocol for the anonymous distribution of qubits \cite{Christandl2005}. This protocol focuses on the use of EPR pairs to transmit a quantum coin via teleportation. Unlike the classical protocol, it doesn't require all honest players to possess the same qubit at the end, avoiding conflicts with the non-cloning property of quantum states. A few years later, Brassard et al. presented their own solution \cite{Brassard2007}. According to their proposal, the sender can transmit a quantum message with complete anonymity even if some participants are corrupted. They introduced the concept of fail-safe teleportation, ensuring that, in the case of quantum teleportation, the information reaches its destination with the highest possible precision and security, even in the presence of errors or disturbances. Later on, Ronghua Shi, Qian Su, Ying Guo, and Moon Ho Lee presented a method for implementing quantum anonymous communication in a public receiver model using the anonymity features of DC-Nets and non-maximally entangled quantum channels \cite{Shi2011}.
Wang and K Zhang analyzed the shortcomings of the above protocols and proposed some modifications, emphasizing the risk to the sender's anonymity in the case of a malicious participant \cite{Wang2014}. In 2015, Ramij Rahaman and Guruprasad Kar presented two quantum protocols addressing the Dining Cryptographers problem and the Anonymous Veto (AV) problem. These protocols are based on the GHZ paradox and the properties of GHZ correlations \cite{Rahaman2015}.
Later, Hameedi et al. proposed a new quantum solution to this problem using a one-way sequential protocol and extending it to the Anonymous Veto problem \cite{Hameedi2017}. The protocol is characterized by relying on a single qubit, utilizing GHZ states due to their high symmetry.
Subsequently, Jan Bouda and Josef Sprojcar presented a protocol for anonymous quantum data sharing \cite{Bouda2007}. According to them, this protocol can implement anonymous sender and receiver channels, ensuring message privacy and anonymity.
In 2021 Li et al. presented an anonymous transmission protocol using single-particle states with collective detection \cite{Li2021}.
Finally, in 2022, Mishra et al. published a series of QAV protocols, quantum protocols for the Anonymous Veto \cite{Mishra2022}.

In this work, we make the case for an innovative entanglement-based protocol for the Dining Cryptographers Problem. The protocol is described as a quantum game, involving signature characters such as Alice, Bob, etc. The pedagogical nature of games is expected to enhance the comprehension of technical concepts. Quantum games, in particular, were initially introduced in 1999 \cite{Meyer1999, Eisert1999}, and have by now gained widespread acceptance. In the realm of quantum cryptography, the presentation of protocols often takes the form of games, a common practice evident in recent works such as \cite{Ampatzis2021, Ampatzis2022, Ampatzis2023, Andronikos2023, Andronikos2023a, Andronikos2023b}, and references therein. Quantum strategies have demonstrated superiority over classical ones in various scenarios \cite{Andronikos2018, Andronikos2021, Andronikos2022a}. The prisoners' dilemma game serves as a prominent example, and its applicability extends to other abstract quantum games \cite{Eisert1999, Giannakis2019}. Notably, the quantization of classical systems, explored in \cite{Andronikos2022}, even finds applications in political structures. In the broader context of game-theoretic applications, unconventional environments, such as biological systems, have garnered significant attention \cite{Theocharopoulou2019, Kastampolidou2020a, Kostadimas2021}. It's intriguing to note that biological systems may give rise to biostrategies that outperform classical ones, even in iconic games like the Prisoners' Dilemma \cite{Kastampolidou2020, Kastampolidou2021, Kastampolidou2023, Papalitsas2021, Adam2023}.

\textbf{Contribution}. This research introduces an innovative entanglement-based protocol to solve the Dining Cryptographers problem, leveraging maximally entangled $\ket{ GHZ_{ n } }$ tuples as its foundation. The primary motivation behind this protocol is to provide scalability in terms of both the number of cryptographers $n$ and the volume of anonymous information conveyed, represented by the number $m$ of qubits within each quantum register.

The protocol accommodates an arbitrary number of cryptographers $n$, allowing scalability not only in participant count but also in the amount of anonymous information transmitted. While the original Dining Cryptographers Problem dealt with a single bit of information—whether a cryptographer paid for dinner—the proposed SQDCP protocol enables $m$, the number of qubits in each register, to be any arbitrarily large positive integer. This flexibility allows for the transmission of diverse information, such as the cost of the dinner or the timing of the arrangement.

A notable aspect of the introduced protocol is its adaptability to both localized and distributed versions of the Dining Cryptographers Problem. The localized scenario involves all cryptographers physically gathering at the same spatial location, like a restaurant, simultaneously. In contrast, the distributed scenario accommodates cryptographers located in different places, engaging in a virtual dinner at the same time.

In terms of implementation, the protocol ensures uniformity as all cryptographers employ identical private quantum circuits. This design establishes a completely modular quantum system where all modules are identical. Additionally, each private quantum circuit exclusively employs the widely-used Hadamard and CNOT quantum gates, facilitating straightforward implementation on contemporary quantum computers.

\section{Background notions} \label{sec: Background Notions}

Quantum physics reveals some astonishing and counterintuitive features that go beyond the limits of classical physics and common sense. One of these amazing phenomena is entanglement, which not only puzzles us but also offers great opportunities for achieving tasks that are hard or even impossible in the classical realm. Entanglement occurs in composite quantum systems, which usually have at least two subsystems, often located at different places. Mathematically, a composite system is said to be entangled when its state cannot be written as a simple product of two or more states of its subsystems. One of the remarkable benefits of quantum entanglement is that when one qubit of an entangled pair or tuple is measured, the other qubit(s) immediately collapse(s) to the matching basis state in the product, no matter how far apart they are. This is exactly the reason why quantum entanglement is useful for various quantum cryptographic protocols, such as key distribution and secret sharing, among others.

Probably, the most famous examples of maximal entanglement are pairs of qubits in one of the four Bell states, also known as EPR pairs. For more information, including their exact mathematical formulation, the curious reader may refer to any standard textbook, such as \cite{Nielsen2010, Yanofsky2013a, Wong2022}. Luckily, maximal entanglement can be easily and intuitively extended to the case of multipartite systems. Perhaps, the most well-known form of maximal entanglement found in composite systems of $n$ qubits, where $n \geq 3$, is the $\ket{ GHZ_{ n } }$ state. The acronym GHZ comes from the initials of the authors Greenberger, Horne, and Zeilinger. In such a case, a composite quantum system is made of $n$ separate qubits, where each subsystem is just a single qubit. The astonishing fact is that the entangled qubits can very well be spatially separated, something to leads to intriguing possibilities, such as quantum teleportation and superdense coding \cite{Ghosh2002, Muralidharan2008}. The $\ket{ GHZ_{ n } }$ state is a maximally entangled state, meaning that the entanglement between the n qubits is as strong as it can be. Mathematically, the situation can be expressed as:

\begin{align} \label{eq: Extended General GHZ_n State}
	\ket{ GHZ_{ n } }
	=
	\frac
	{
		\ket{ 0 }_{ n - 1 } \ket{ 0 }_{ n - 2 } \dots \ket{ 0 }_{ 0 }
		+
		\ket{ 1 }_{ n - 1 } \ket{ 1 }_{ n - 2 } \dots \ket{ 1 }_{ 0 }
	}
	{ \sqrt{ 2 } }
	\ ,
\end{align}

where the subscript $i, \ 0 \leq i \leq n - 1$, denotes the $i^{ th }$ qubit.

Currently, contemporary quantum computers can generate arbitrary GHZ states utilizing standard quantum gates like the Hadamard and CNOT gates. Furthermore, the circuits responsible for producing these states exhibit high efficiency, requiring only $\lg n$ steps for the $\ket{ GHZ_{ n } }$ state, as demonstrated in \cite{Cruz2019}.

The full power of the introduced protocol requires a more complex and versatile quantum system where each subsystem is a quantum register $r_{ i }$, $0 \leq i \leq n - 1$, containing $m$ qubits. The distinguishing characteristic of this setting is that corresponding qubits across all $n$ registers are entangled in the $\ket{ GHZ_{ n } }$ state. This concept is formally captured by the Symmetric Bitwise Entanglement Distribution Scheme, which is employed by the current protocol. The next Definition \ref{def: Entanglement Distribution Scheme} provides the details.

\begin{definition} \label{def: Entanglement Distribution Scheme}
	The $( n, m )$ Symmetric Bitwise Entanglement Distribution Scheme, denoted by $SBEDS_{ n, m }$, specifies that:
	\begin{itemize}
		\item[$\blacktriangleright$]	there are $n$ quantum registers $r_0, r_{ 1 }, \dots, r_{ n - 1 }$,
		\item[$\blacktriangleright$]	each register contains $m$ qubits, and
		\item[$\blacktriangleright$]	the $n$ qubits in position $j, \ 0 \leq j \leq m - 1$, in every register are entangled in the $\ket{ GHZ_{ n } }$ state.
	\end{itemize}
	It is important to clarify that the system of the $n$ quantum registers can be either local, or spatially distributed, depending on whether or not the registers are in the same or different geographical locations in space.
\end{definition}

In any event, the global state of the composite system is given by the following equation.

\begin{align} \label{eq: m-Fold Extended General GHZ_n State}
	\ket{ GHZ_{ n } }^{ \otimes m }
	&=
	\frac { 1 } { \sqrt{ 2^{ m } } }
	\sum_{ \mathbf{ x } \in \mathbb{ B }^{ m } }
	\ket{ \mathbf{ x } }_{ n - 1 } \dots \ket{ \mathbf{ x } }_{ 0 }
	\ .
\end{align}

In writing formula \eqref{eq: m-Fold Extended General GHZ_n State}, which is proved in \cite{Ampatzis2023}, we take advantage of the standard notation explained below.

\begin{itemize}
	\item	$\mathbb{ B }$ is the binary set $\{ 0, 1 \}$.
	\item	To distinguish bit vectors from bits, we follow the accepted convention of writing bit vectors $\mathbf{ x } \in \mathbb{ B }^{ m }$ in boldface. A bit vector $\mathbf{ x }$ of length $m$ is an abbreviation of a sequence of $m$ bits: $\mathbf{ x } = x_{ m - 1 } \dots x_{ 0 }$. In the special case where all bits are zero, i.e., $0 \dots 0$, we speak of the zero bit vector, which is denoted by $\mathbf{ 0 }$.
	\item	In this setting, a bit vector $\mathbf{ x } \in \mathbb{ B }^{ m }$ stands for the binary representation of one of the $2^{ m }$ basis kets that form the computational basis of the Hilbert space at hand.
	\item	To eliminate any source of ambiguity, we rely on the indices $i, \ 0 \leq i \leq n - 1$, to emphasize that $\ket{ \mathbf{ x } }_{ i }$ is the state of the $i^{ th }$ quantum register.
\end{itemize}

Figure \ref{fig: The Entanglement Distribution Scheme} vividly depicts this setting, with the corresponding qubits that form the $\ket{ GHZ_{ n } }$ $n$-tuple colored identically. This composite system comprises $m n$ qubits as each of the $n$ registers holds $m$ qubits. We point out that it doesn't matter in the least if the registers are all in the same place, or are all in different spatial locations. The power of the entanglement effect, stemming from the $m$ $\ket{ GHZ_{ n } }$ $n$-tuples, will instill the necessary correlation, irrespective of whether the composite system is localized or entirely distributed. It is precisely this unique effect of entanglement that allows us to envision the whole setting as a unified system.

Apart from $\ket{ GHZ_{ n } }$ tuples that are necessary for the operation of our protocol, another hallmark state, namely $\ket{ - }$, features prominently.

	\begin{align} \label{eq: Ket -}
		\ket{ - } = H \ket{ 1 } = \frac { \ket{ 0 } - \ket{ 1 } } { \sqrt{ 2 } }
	\end{align}

The forthcoming mathematical analysis will also use the important formula that expresses the $m$-fold Hadamard transform of an arbitrary basis ket shown below. Its proof is omitted because it can be easily found in most standard textbooks, e.g., \cite{Nielsen2010, Mermin2007}.

\begin{align} \label{eq: Hadamard m-Fold Ket x}
	H^{ \otimes m } \ket{ \mathbf{ x } }
	&=
	\frac { 1 } { \sqrt{ 2^{ n } } }
	\sum_{ \mathbf{ z } \in \mathbb{ B }^{ m } }
	( - 1 )^{ \mathbf{ z \bullet x } } \ket{ \mathbf{ z } }
	\ .
\end{align}

\begin{tcolorbox}
	[
		grow to left by = 0.00 cm,
		grow to right by = 0.00 cm,
		colback = MagentaLight!03,			
		enhanced jigsaw,					
		sharp corners,
		toprule = 1.0 pt,
		bottomrule = 1.0 pt,
		leftrule = 0.1 pt,
		rightrule = 0.1 pt,
		sharp corners,
		center title,
		fonttitle = \bfseries
	]
	\begin{figure}[H]
		\centering
		\begin{tikzpicture} [ scale = 0.750 ]
			\node
				[
					anchor = center,
				]
				(Alice) { $\mathbf{ r_{ n - 1 } }$: };
			\matrix
				[
					matrix of nodes, nodes in empty cells,
					column sep = 3.00 mm, right = 0.50 of Alice,
					nodes = { circle, minimum size = 12 mm, semithick, font = \footnotesize },
				]
				{
					\node [ shade, outer color = RedPurple!50, inner color = white ] (A_m-1^n-1) { $q_{ m - 1 }^{ n - 1 }$ }; &
					\node [ shade, outer color = WordAquaLighter60, inner color = white ] { \large \dots }; &
					\node [ shade, outer color = GreenLighter2!50, inner color = white ] (A_1^n-1) { $q_{ 1 }^{ n - 1 }$ }; &
					\node [ shade, outer color = WordBlueVeryLight, inner color = white ] (A_0^n-1) { $q_{ 0 }^{ n - 1 }$ };
					\\
				};
			\node
				[
					anchor = center,
					above = 1.50 cm of Alice,
				]
				(Bob) { $\mathbf{ r_{ n - 2 } }$: };
			\matrix
				[
					matrix of nodes, nodes in empty cells,
					column sep = 3.00 mm, right = 0.50 of Bob,
					nodes = { circle, minimum size = 12 mm, semithick, font = \footnotesize },
				]
				{
					\node [ shade, outer color = RedPurple!50, inner color = white ] (B_m-1^n-2) { $q_{ m - 1 }^{ n - 2 }$ }; &
					\node [ shade, outer color = WordAquaLighter60, inner color = white ] { \large \dots }; &
					\node [ shade, outer color = GreenLighter2!50, inner color = white ] (B_1^n-2) { $q_{ 1 }^{ n - 2 }$ }; &
					\node [ shade, outer color = WordBlueVeryLight, inner color = white ] (B_0^n_2) { $q_{ 0 }^{ n - 2 }$ };
					\\
				};
			\node
				[
					anchor = center,
					above = 1.00 cm of Bob,
				]
				(Dots) { \Large \vdots };
			\matrix
				[
					right = 3.75 cm of Dots, anchor = center, column sep = 0.00 mm, row sep = 0.0 mm,
					nodes = { minimum size = 14 mm, semithick }
				]
				{
					\node { }; &
					\node { }; &
					\node { \Large \vdots }; &
					\node { }; &
					\node { };
					\\
				};
			\node
				[
					anchor = center,
					above = 0.80 cm of Dots,
				]
				(Charlie) { $\mathbf{ r_{ 1 } }$: };
			\matrix
				[
					matrix of nodes, nodes in empty cells,
					column sep = 3.00 mm, right = 0.70 of Charlie,
					nodes = { circle, minimum size = 12 mm, semithick, font = \footnotesize },
				]
				{
					\node [ shade, outer color = RedPurple!50, inner color = white ] (C_m-1^1) { $q_{ m - 1 }^{ 1 }$ }; &
					\node [ shade, outer color = WordAquaLighter60, inner color = white ] { \large \dots }; &
					\node [ shade, outer color = GreenLighter2!50, inner color = white ] (C_1^1) { $q_{ 1 }^{ 1 }$ }; &
					\node [ shade, outer color = WordBlueVeryLight, inner color = white ] (C_0^1) { $q_{ 0 }^{ 1 }$ };
					\\
				};
			\node
				[
					anchor = center,
					above = 1.50 cm of Charlie,
				]
				(Dave) { $\mathbf{ r_{ 0 } }$: };
			\matrix
				[
					matrix of nodes, nodes in empty cells,
					column sep = 3.00 mm, right = 0.70 of Dave,
					nodes = { circle, minimum size = 12 mm, semithick, font = \footnotesize },
				]
				{
					\node [ shade, outer color = RedPurple!50, inner color = white ] (D_m-1^0) { $q_{ m - 1 }^{ 0 }$ }; &
					\node [ shade, outer color = WordAquaLighter60, inner color = white ] { \large \dots }; &
					\node [ shade, outer color = GreenLighter2!50, inner color = white ] (D_1^0) { $q_{ 1 }^{ 0 }$ }; &
					\node [ shade, outer color = WordBlueVeryLight, inner color = white ] (D_0^0) { $q_{ 0 }^{ 0 }$ };
					\\
				};
			\node
				[
					above right = 3.50 cm and 3.25 cm of Dave, anchor = center,
					anchor = center,
					shade,
					top color = GreenTeal, bottom color = black,
					rectangle,
					text width = 12.00 cm,
					align = center
				]
				(Label)
				{ \color{white} \textbf{A composite system of $n$ quantum registers $r_{ 0 }, \dots, r_{ n - 1 }$, each with $m$ qubits. The characteristic property of this system is that the qubits in the corresponding positions make up a $\ket{ GHZ_{ n } }$ $n$-tuple.} };
			\begin{scope}[on background layer]
				\node
					[
						above = - 1.50 cm of A_m-1^n-1,
						rectangle,
						rounded corners = 8 pt,
						fill = RedPurple!15,
						minimum width = 6 mm,
						minimum height = 88 mm
					]
					( ) { };
				\node
					[
						above = 0.75 cm of D_m-1^0
					]
					( ) {\colorbox{RedPurple!50} {$\ket{ GHZ_{ n } }$}};
				\node
					[
						above = - 1.50 cm of A_1^n-1,
						rectangle,
						rounded corners = 8 pt,
						fill = GreenLighter2!15,
						minimum width = 6 mm,
						minimum height = 88 mm
					]
					( ) { };
				\node
					[
						above = 0.75 cm of D_1^0
					]
					( ) {\colorbox{GreenLighter2!50} {$\ket{ GHZ_{ n } }$}};
				\node
					[
						above = - 1.50 cm of A_0^n-1,
						rectangle,
						rounded corners = 8 pt,
						fill = WordBlueVeryLight!30,
						minimum width = 6 mm,
						minimum height = 88 mm
					]
					( ) { };
				\node
					[
						above = 0.75 cm of D_0^0
					]
					( ) {\colorbox{WordBlueVeryLight} {$\ket{ GHZ_{ n } }$}};
			\end{scope}
			\node
				[
					anchor = east,
					below = 1.00 cm of Alice
				]
				(PhantomNode) { };
		\end{tikzpicture}
		\caption{This figure draws the $n$ qubits that populate the same position in the $r_{ 0 }, \dots, r_{ n - 1 }$ registers with the same color. This is done to drive the point that they belong to the same $\ket{ GHZ_{ n } }$ $n$-tuple.}
		\label{fig: The Entanglement Distribution Scheme}
	\end{figure}
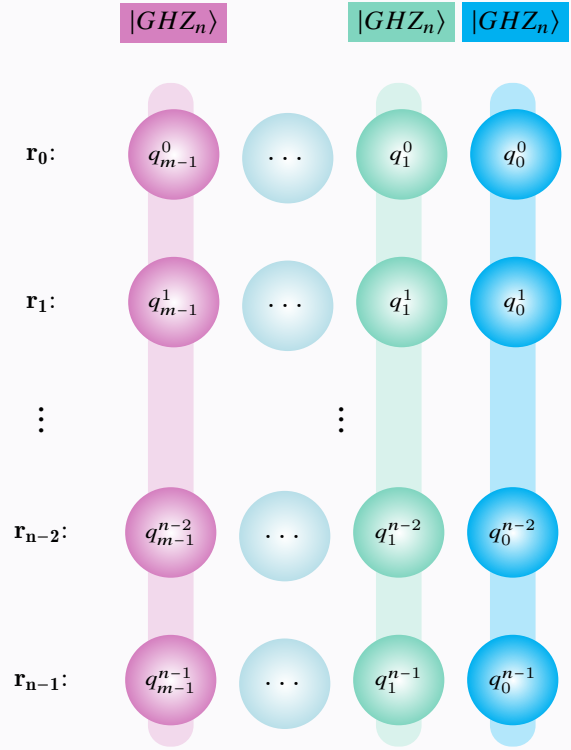
\end{tcolorbox}

In the previous relation \eqref{eq: Hadamard m-Fold Ket x}, the expression $\mathbf{ x \bullet y }$ denotes the \emph{inner product modulo} $2$ operation. The inner product modulo $2$ takes as inputs two bit vectors $\mathbf{ x }, \mathbf{ y } \in \mathbb{ B }^{ m }$, and outputs their inner product. Specifically, if $\mathbf{ x } = x_{ m - 1 } \dots x_{ 0 }$ and $\mathbf{ y } = y_{ m - 1 } \dots y_{ 0 }$, then $\mathbf{ x \bullet y }$ is computed as

\begin{align} \label{eq: Inner Product Modulo $2$}
	\mathbf{ x \bullet y }
	&=
	x_{ m - 1 } y_{ m - 1 }
	\oplus \dots \oplus
	x_{ 0 } y_{ 0 }
	\ ,
\end{align}

where $\oplus$ is \emph{addition modulo} $2$. The inner product modulo $2$ operation is characterized by a fundamental property whose application is central to many quantum algorithms. Consider any fixed element $\mathbf{ c }$ of $\mathbb{ B }^{ m }$; if $\mathbf{ c }$ is different from $\mathbf{ 0 }$, then for half of the elements $\mathbf{ x } \in \mathbb{ B }^{ m }$, the outcome of the operation $\mathbf{c} \bullet \mathbf{ x }$ is $0$, and for the remaining half, the result of the operation $\mathbf{ c } \bullet \mathbf{ x }$ is $1$. The exception of $\mathbf{ 0 }$ is justified because when $\mathbf{ c } = \mathbf{ 0 }$, then for all $\mathbf{ x } \in \mathbb{ B }^{ m }$, $\mathbf{ c } \bullet \mathbf{ x } = 0$. Following \cite{Andronikos2023b}, we call this property the characteristic inner product (CIP) property.

		\begin{align}
			\mathbf{ c } = \mathbf{ 0 }
			&\Rightarrow
			\text{for all } 2^{ m } \text{ bit vectors } \mathbf{ x } \in \mathbb{ B }^{ m },
			\text{ } \mathbf{ c } \bullet \mathbf{ x } = 0
			\label{eq: Inner Product Modulo $2$ Property For Zero}
			\\
			\mathbf{ c } \neq \mathbf{ 0 }
			&\Rightarrow
			\left\{
			\
			\begin{matrix*}[l]
				\text{for } 2^{ m - 1 } \text{ bit vectors } \mathbf{ x } \in \mathbb{ B }^{ m }, \ \mathbf{ c } \bullet \mathbf{ x } = 0
				\\
				\text{for } 2^{ m - 1 } \text{ bit vectors } \mathbf{ x } \in \mathbb{ B }^{ m }, \ \mathbf{ c } \bullet \mathbf{ x } = 1
			\end{matrix*}
			\
			\right\}
			\label{eq: Inner Product Modulo $2$ Property For NonZero}
		\end{align}

For completeness, we also clarify that measurements are performed with respect to the computational basis $\{ \ket{ 0 }, \ket{ 1 } \}$.
\section{The Scalable quantum protocol for the Dinning Cryptographers Problem} \label{sec: The SQDCP Protocol}

In the current section, we introduce the novel entanglement-based scalable protocol that solves the Dining Cryptographers Problem, or SQDCP protocol for short. As is often done with most of the cryptographic protocols, to lighten up the presentation and make it more easy-going, we employ the format of a quantum game. This game is played by $n$ cryptographers, one of which is Alice, the star of the game. Alice has organized an official dinner for herself and her $n - 1$ colleagues $C_{ 0 }$, \dots, $C_{ n - 2 }$. In the forthcoming small-scale examples, the roles of Alice's colleagues will be assumed by other famous actors, namely Bod, Charlie, and Dave. Nonetheless, in the formal presentation of the protocol, and in order to stress its scalability for arbitrarily large values of $n$, the $n - 1$ cryptographers will not be given particular names. Our $n$ heroes, upon discovering that the dinner has been paid, set out to discover whether it was one of them or their employer that paid the bill. The crucial thing here is that they must find out the truth in such a way so that the anonymity and privacy of all cryptographers is not compromised in any way whatsoever and the identity of the cryptographer who paid remains unknown. In its essence, the Dinning Cryptographers Problem is a metaphor for anonymous and untraceable information transmission.

As we mentioned in the Introduction, the SQDCP protocol brings three main novelties to the table. So, before we begin its description in earnest, let us explain them in detail.

\begin{enumerate} [ left = 0.50 cm, labelsep = 1.00 cm, start = 1 ]
	\renewcommand\labelenumi{(\textbf{N}$_{ \theenumi }$)}
	\item	\textbf{Scalability}. In the SQDCP protocol, the notion of scalability encompasses both parameters $n$ and $m$. The number of cryptographers $n$ can be any arbitrary large positive integer. In addition to the scalability of players, our protocol can seamlessly scale in terms of the amount of anonymous information it conveys. Initially, the Dinning Cryptographers Problem was about just one bit of information, namely if one of the cryptographers paid or not for the dinner. In the SQDCP protocol, the number $m$ of qubits in each register can also be any large positive integer. This number reflects the amount of information that can be transmitted. For instance, the cryptographer that actually paid the bill, may also disclose how much it cost or when the arrangement was made, etc.
	\item	\textbf{Local and Distribute Mode}. In its initial formulation in \cite{Chaum1988} and in the subsequent treatments, the cryptographers' dinner was a localized event, in the sense that all cryptographers were together at the same spatial location at a specific point in time. The protocol introduced in this work can address not only this localized situation but also a distributed version of the Dining Cryptographers Problem, in which the cryptographers are in different spatial locations.
	\item	\textbf{Uniformity and Ease of Implementation}. All cryptographers employ identical quantum circuits, that is the quantum implementation is completely modular, with all modules being the same. Furthermore, each quantum circuit can be easily implemented by a contemporary quantum computer because it only uses the ubiquitous Hadamard and CNOT quantum gates.
\end{enumerate}

Given the above considerations, the following two definitions state formally the localized and the distributed versions of the Dining Cryptographers Problem.

\begin{definition} [Localized Setting] \label{def: The Localized DCP}
	The localized setup is described below.
	\begin{itemize}
		\item	Alice gathers together with her $n - 1$ cryptographer colleagues $C_{ 0 }$, \dots, $C_{ n - 2 }$ for a friendly dinner in a nearby restaurant.
		\item	For all $n$ players, the dinner event takes place simultaneously and at the same location.
		\item	Each player employs a quantum circuit where she secretly embeds the desired information, namely whether she paid or not for the dinner.
		\item	Upon measuring their quantum registers and publicly combining the obtained results, all the players immediately know if the dinner was paid by one of them or by their employer, and, possibly, some additional information, e.g., the cost of the dinner or the date of the payment, etc.
		\item	The identity of the one who paid the bill must remain unknown to all other cryptographers.
	\end{itemize}
\end{definition}

One of the most useful traits of entanglement is that entangled subsystems are intertwined despite being spatially separated. This is the key that enables the effective operation of many distributed quantum protocols. The SQDCP protocol also takes advantage of entanglement in order to achieve the desired outcome in a completely distributed setting.

\begin{definition} [Distributed Setting] \label{def: The Distributed DCP}
	Let us now envision a more general situation.
	\begin{itemize}
		\item	Alice and her $n - 1$ cryptographer colleagues $C_{ 0 }$, \dots, $C_{ n - 2 }$ have made arrangement for dinner.
		\item	There is a complication now compared to the previous case because all $n$ agents reside at different geographical locations.
		\item	Nevertheless, they are determined to dine at the same time, albeit in different restaurants, and be in constant audio and visual contact via teleconference.
		\item	Each player employs a local quantum circuit where she secretly embeds the desired information, namely whether she paid or not for the dinner.
		\item	Upon measuring their quantum registers, they publicly exchange their measurements via classical channels. Subsequently, each player combines the received results and immediately knows if the dinner was paid by one of them or by their employer, and, possibly, some additional information.
		\item	The identity of the one who paid the bill must remain unknown to all other cryptographers.
	\end{itemize}
\end{definition}

The task at hand is to devise a quantum protocol that can seamlessly operate in both localized and distributed modes, and reveal the required information while guaranteeing the privacy and anonymity of the generous cryptographer.

Before we proceed with the detailed description of the protocol, it will be expedient to make some clarifications, to avoid any possible confusion.

\begin{itemize}
	\item	Although there is no theoretical limitation on the number $n$ of cryptographers that can be an arbitrarily large integer, contemporary quantum apparatus may impose constraints to the generation of $\ket{ GHZ_{ n } }$ tuples, whenever $n$ exceeds some threshold.
	\item	We assume that, prior to the execution of the SQDCP protocol, certain arrangements have taken place among the cryptographers regarding the amount and nature of the desired information. This is necessary in order to fix the number of $m$, corresponding to the amount of information, and the proper interpretation of the outcome.
	\item	In the distributed version, we also assume the existence of pairwise authenticated channels that enable the transmission of classical information.
\end{itemize}

\begin{example} \label{xmp: Alice, Bob, Charlie & Dave's Dinner} \
	This example features the four cryptographers Alice, Bob, Charlie, and Dave having dinner in both the localized and the distributed scenario. Common to both scenarios is the fact that their input quantum registers are entangled according to the $( n, m )$ Symmetric Bitwise Entanglement Distribution Scheme, explained in Definition \ref{def: Entanglement Distribution Scheme}. In this special case, $n = 4$ and, if we assume that the desired information includes also the amount of money paid for the dinner, we may also take $m$ equal to $4$. Hence, the final outcome of the protocol will be $m$ bits expressing the binary representation of the cost of the dinner, say in euros \texteuro. This small-scale example requires $4$ $\ket{ GHZ_{ 4 } }$ evenly and uniformly distributed among the $4$ players.

	When the dinner event takes place at the same restaurant at the same time for all four cryptographers, we have the standard version of the Dining Cryptographers Problem. This is what we refer to as the localized scenario, which is illustrated in Figure \ref{fig: The Localized DCP}.

\begin{tcolorbox}
	[
		grow to left by = 0.00 cm,
		grow to right by = 0.00 cm,
		colback = MagentaLight!03,			
		enhanced jigsaw,					
		sharp corners,
		toprule = 1.0 pt,
		bottomrule = 1.0 pt,
		leftrule = 0.1 pt,
		rightrule = 0.1 pt,
		sharp corners,
		center title,
		fonttitle = \bfseries
	]
	\begin{figure}[H]
		\centering
		\begin{tikzpicture} [scale = 1.00]
			\def \n {4}
			\def \Angle {360 / \n}
			\def \Radius {3.00}
			\scoped [ on background layer ]
				\shadedraw [ inner color = orange!50!black, outer color = brown, draw = brown ] ( 0, 0 ) circle [ radius = \Radius cm ];
			\node
				[
					dave,
					scale = 2.50,
					anchor = center,
					label = { [ label distance = 0.00 cm ] south: Dave }
				]
				(Dave) at ( { 1.25 * \Radius * cos( 3 * \Angle ) }, { 1.25 * \Radius * sin( 3 * \Angle ) } ) { };
			\node [ scale = 2.0, above = 1.00 cm of Dave, anchor = center ] { \faUtensils };
			\node
				[
					charlie,
					scale = 2.50,
					anchor = center,
					label = { [ label distance = 0.00 cm ] west: Charlie }
				]
				(Charlie) at ( { 1.2 * \Radius * cos( 2 * \Angle ) }, { 1.2 * \Radius * sin( 2 * \Angle ) } ) { };
			\node [ scale = 2.0, right = 1.00 cm of Charlie, anchor = center ] { \faUtensils };
			\node
				[
					bob,
					scale = 2.50,
					anchor = center,
					label = { [ label distance = 0.00 cm ] north: Bob }
				]
				(Bob) at ( { 1.25 * \Radius * cos( 1 * \Angle ) }, { 1.25 * \Radius * sin( 1 * \Angle ) } ) { };
			\node [ scale = 2.0, below = 1.00 cm of Bob, anchor = center ] { \faUtensils };
			\node
				[
					alice,
					scale = 2.50,
					anchor = center,
					label = { [ label distance = 0.00 cm ] east: Alice }
				]
				(Alice) at ( { 1.2 * \Radius * cos( 0 * \Angle ) }, { 1.2 * \Radius * sin( 0 * \Angle ) } ) { };
			\node [ scale = 2.0, left = 1.00 cm of Alice, anchor = center ] { \faUtensils };
			\node
				[
					above = 2.00 cm of Bob, anchor = center,
					anchor = center,
					shade,
					top color = GreenTeal, bottom color = black,
					rectangle,
					text width = 12.00 cm,
					align = center
				]
				(Label)
				{ \color{white} \textbf{\underline{The Localized Scenario}}
				\\
				The cryptographers Alice, Bob, Charlie \& Dave have gathered together at a restaurant for dinner. They want to find out if one of them has paid for this dinner, but without disclosing her or his identity.};
		\end{tikzpicture}
		\caption{The above figure visualizes an example of the localized scenario. Four cryptographers, Alice, Bob, Charlie \& Dave, have gathered together at a restaurant to enjoy their dinner. They want to find out if one of them has paid for this dinner, but without disclosing her or his identity.} \label{fig: The Localized DCP}
	\end{figure}
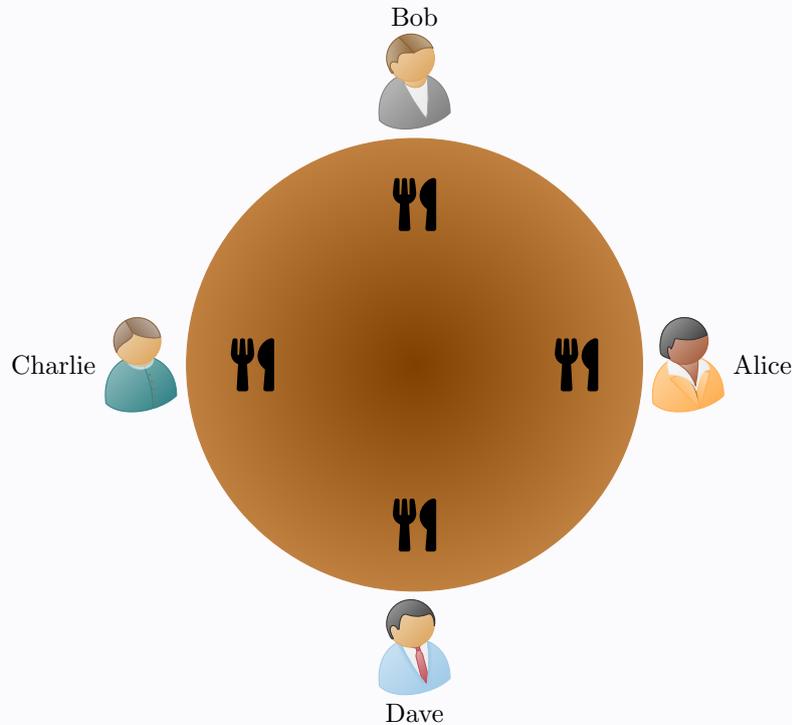
\end{tcolorbox}

	In addition to the standard approach, the fact that the quantum registers are entangled opens up the possibility of implementing the SQDCP protocol even in the case where the cryptographers are spatially separated. In such a scenario, they are having a virtual dinner at the same time, but now each of them resides in a different location, as depicted in Figure \ref{fig: The Distributed DCP}. \hfill $\triangleleft$
\end{example}

\begin{tcolorbox}
	[
		grow to left by = 0.00 cm,
		grow to right by = 0.00 cm,
		colback = MagentaLight!03,			
		enhanced jigsaw,					
		sharp corners,
		toprule = 1.0 pt,
		bottomrule = 1.0 pt,
		leftrule = 0.1 pt,
		rightrule = 0.1 pt,
		sharp corners,
		center title,
		fonttitle = \bfseries
	]
	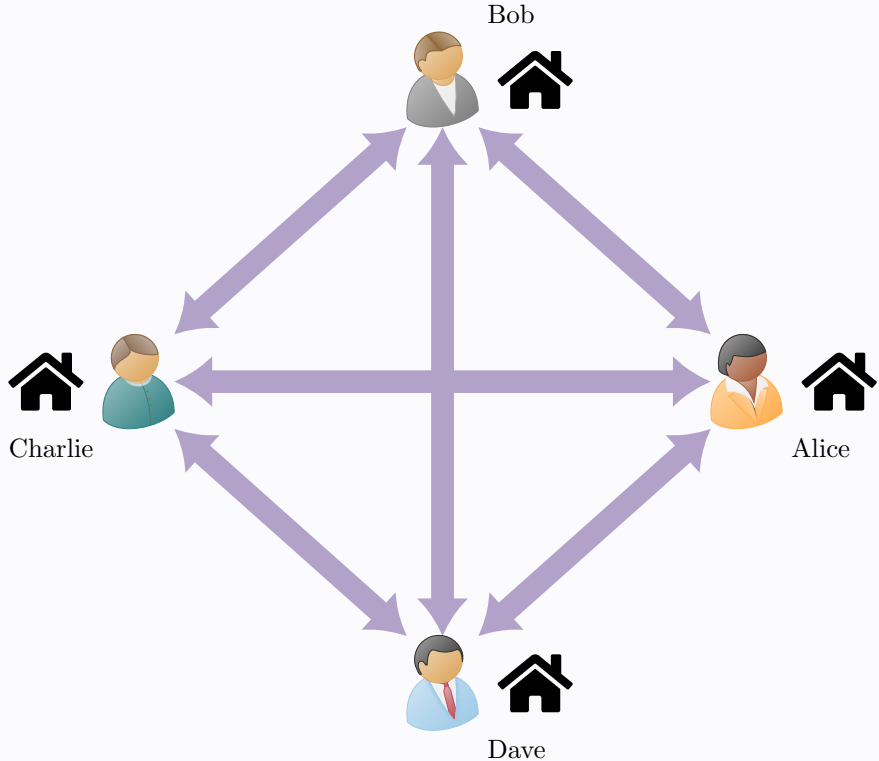
\begin{figure}[H]
		\centering
		\begin{tikzpicture} [scale = 1.00]
			\def \n {8}
			\def \Angle {360 / \n}
			\def \Radius {4.00}
			\node
				[
					dave,
					scale = 2.50,
					anchor = center,
					label = { [ label distance = 0.00 cm ] south east: Dave }
				]
				(Dave) at ( { \Radius * cos( 6 * \Angle ) }, { \Radius * sin( 6 * \Angle ) } ) { };
			\node [ scale = 2.5, right = 0.75 cm of Dave, anchor = center ] { \faHome };
			\node
				[
					charlie,
					scale = 2.50,
					anchor = center,
					label = { [ label distance = 0.00 cm ] south west: Charlie }
				]
				(Charlie) at ( { \Radius * cos( 4 * \Angle ) }, { \Radius * sin( 4 * \Angle ) } ) { };
			\node [ scale = 2.5, left = 0.75 cm of Charlie, anchor = center ] { \faHome };
			\node
				[
					bob,
					scale = 2.50,
					anchor = center,
					label = { [ label distance = 0.00 cm ] north east: Bob }
				]
				(Bob) at ( { \Radius * cos( 2 * \Angle ) }, { \Radius * sin( 2 * \Angle ) } ) { };
			\node [ scale = 2.5 , right = 0.75 cm of Bob, anchor = center ] { \faHome };
			\node
				[
					alice,
					scale = 2.50,
					anchor = center,
					label = { [ label distance = 0.00 cm ] south east: Alice }
				]
				(Alice) at ( { \Radius * cos( 0 * \Angle ) }, { \Radius * sin( 0 * \Angle ) } ) { };
			\node [ scale = 2.5, right = 0.75 cm of Alice, anchor = center ] { \faHome };
			\begin{scope}[on background layer]
				\draw [ MagentaVeryLight, <->, { Latex [ width = 7mm, length = 5mm ] }-{ Latex [ width = 7mm, length = 5mm ] }, line width = 3.00 mm ]
				(Alice.north west) -- (Bob.south east);
				\draw [ MagentaVeryLight, <->, { Latex [ width = 7mm, length = 5mm ] }-{ Latex [ width = 7mm, length = 5mm ] }, line width = 3.00 mm ]
				(Alice.west) -- (Charlie.east);
				\draw [ MagentaVeryLight, <->, { Latex [ width = 7mm, length = 5mm ] }-{ Latex [ width = 7mm, length = 5mm ] }, line width = 3.00 mm ]
				(Alice.south west) -- (Dave.north east);
				\draw [ MagentaVeryLight, <->, { Latex [ width = 7mm, length = 5mm ] }-{ Latex [ width = 7mm, length = 5mm ] }, line width = 3.00 mm ]
				(Bob.south west) -- (Charlie.north east);
				\draw [ MagentaVeryLight, <->, { Latex [ width = 7mm, length = 5mm ] }-{ Latex [ width = 7mm, length = 5mm ] }, line width = 3.00 mm ]
				(Bob.south) -- (Dave.north);
				\draw [ MagentaVeryLight, <->, { Latex [ width = 7mm, length = 5mm ] }-{ Latex [ width = 7mm, length = 5mm ] }, line width = 3.00 mm ]
				(Charlie.south east) -- (Dave.north west);
			\end{scope}
			\node
				[
					above = 2.00 cm of Bob, anchor = center,
					anchor = center,
					shade,
					top color = GreenTeal, bottom color = black,
					rectangle,
					text width = 12.00 cm,
					align = center
				]
				(Label)
				{ \color{white} \textbf{\underline{The Distributed Scenario}}
				\\
				The cryptographers Alice, Bob, Charlie \& Dave have arranged a virtual dinner because the reside in different locations. They still want to find out if one of them has paid for this dinner, but without disclosing her or his identity.};
		\end{tikzpicture}
		\caption{The above figure illustrates an example of the distributed scenario. Four cryptographers, Alice, Bob, Charlie \& Dave, have arranged a virtual dinner using state-of-the-art technology because they are in different geographical locations. Of course, they still want to find out if one of them has paid for this dinner, but without disclosing her or his identity.} \label{fig: The Distributed DCP}
	\end{figure}
\end{tcolorbox}

Having explained the general philosophy and intuition behind our protocol, we proceed to its detailed presentation in the next section.

\section{Execution of the SQDCP protocol in $3$ phases} \label{sec: Execution of the SQDCP Protocol in $3$ phases}

For pedagogical reasons, we may conceptually view the protocol as evolving in $3$ phases. Before we begin in earnest, let us point out that our protocol does not violate the no-cloning theorem \cite{wootters1982single}. The state of the distributed $\ket{ GHZ_{ n } }$ tuples is known to all cryptographers participating in the protocol. At a subsequent phase, one of the cryptographers embeds the private information into the distributed entangled state of the system.

\subsection{Entanglement distribution phase} \label{subsec: Entanglement Distribution Phase}

The first phase of the protocol is the entanglement distribution phase, during which the following actions take place. In the subsequent analysis, the number qubits in every input quantum register is designated by $m$. This number is taken to be an appropriately chosen large positive integer, capable of conveying the amount of information desired. In the rest of this article, we mainly consider the case where this information is the amount of money paid for the dinner.

\begin{enumerate} [ left = 0.750 cm, labelsep = 1.00 cm ]
	\renewcommand\labelenumi{(\textbf{ED}$_\theenumi$)}
	\item	Alice or perhaps a third party, trusted by all cryptographers, generates a sequence of $m$ $\ket{ GHZ_{ n } }$ tuples, $m n$ qubits in total, which are necessary for the execution of the protocol and the private transmission of the required information. For the SQDCP protocol, the exact source responsible for the production of the $\ket{ GHZ_{ n } }$ tuples is not important; the only thing that matters is that they are faithfully created and distributed among the cryptographers.
	\item	Say for convenience that in every $\ket{ GHZ_{ n } }$ tuple the qubits are numbered from $0$ to $n - 1$. Their distribution adheres to the following pattern, which guarantees the even and uniform distribution of entanglement among the cryptographers.
	\begin{itemize}
		\item[$\blacktriangleright$]	Alice stores in her input register, denoted by $AIR$ in Figure \ref{fig: The Quantum Circuit for the SQDCP Protocol}, the $( n - 1 )^{ th }$ qubit of each $\ket{ GHZ_{ n } }$ tuple.
		\item[$\blacktriangleright$]	Cryptographer $C_{ i }$, $0 \leq i \leq n - 2$, stores in her input register, symbolized by $IR_{ i }$ in Figure \ref{fig: The Quantum Circuit for the SQDCP Protocol}, the $i^{ th }$ qubit of each $\ket{ GHZ_{ n } }$ tuple.
	\end{itemize}
	\item	In addition to her input register, Alice utilizes a single-bit output register designated by $AOR$ in Figure \ref{fig: The Quantum Circuit for the SQDCP Protocol}, which is initialized at state $\ket{ - }$. Likewise, all her cryptographer colleagues $C_{ i }$, $0 \leq i \leq n - 2$, possess a similar single bit output register denoted by $OR_{ i }$ in Figure \ref{fig: The Quantum Circuit for the SQDCP Protocol}. These output registers play a crucial role in the embedding of private information into the entangled state of the composite circuit.
\end{enumerate}

All cryptographers operate privately and secretly on their own quantum circuits. The important remark in this respect is that all these circuits are identical. In closing this subsection, we point out that the entire algorithm hinges upon the existence of entanglement, and, therefore, it is very important to verify that all the cryptographers' quantum registers are appropriately entangled. Given its significance, this task has undergone thorough scrutiny in the existing literature. Our protocol adheres to the sophisticated methodologies outlined in prior works, including \cite{Neigovzen2008, Feng2019, Wang2022a, Yang2022, Qu2023, Ikeda2023c}. Hence, to preclude redundant exposition, we direct the reader to the previously mentioned bibliography for all the details essential for the successful validation of the entanglement.

\subsection{Private information embedding phase} \label{subsec: Private Information Embedding Phase}

During this phase, the cryptographer who actually paid for the dinner, assuming that this was indeed the case, is able to privately and secretly embed this information, perhaps along with any additional related information. In particular, the quantum circuit outlined in Figure \ref{fig: The Quantum Circuit for the SQDCP Protocol} operates as described below.

\begin{enumerate} [ left = 0.850 cm, labelsep = 1.00 cm ]
	\renewcommand\labelenumi{(\textbf{PIE}$_\theenumi$)}
	\item	If it was Alice who secretly paid for the dinner in question, and the binary representation of the amount she paid is $\mathbf{ p }_{ A }$ in euros \texteuro, then she will insert $\mathbf{ p }_{ A }$ into the global entangled state of the circuit via her private unitary transform $U_{ f_{ A } }$. Since $U_{ f_{ A } }$ is only known to her, the required information will be embedded secretly, privately, and none will be able to be trace it back to Alice.
	\item	If Alice didn't pay for the dinner, then she uses the zero bit vector $\mathbf{ 0 }$ in her private unitary transform $U_{ f_{ A } }$, which in effect leaves the global state of the system unchanged, and doesn't embed any new information.
	\item	Entirely analogously, if it was cryptographer $C_{ i }$, $0 \leq i \leq n - 2$, who secretly paid for the dinner in question, and the binary representation of the amount paid is $\mathbf{ p }_{ i }$, then she will insert $\mathbf{ p }_{ i }$ into the global entangled state of the circuit via her private unitary transform $U_{ f_{ i } }$. Since $U_{ f_{ i } }$ is only known to cryptographer $C_{ i }$, this information will be embedded secretly, and privately, and none will be able to trace it back to $C_{ i }$. Obviously, if $C_{ i }$ didn't pay for the dinner, then she uses the zero bit vector $\mathbf{ 0 }$ in her private unitary transform $U_{ f_{ i } }$, which doesn't encode any new information to the global state of the circuit.
	\item	The quantum part of the protocol is completed when the cryptographers measure their input registers. The obtained measurements are added together using addition modulo $2$, i.e., they are XOR-ed together. The final outcome $\mathbf{ p }$ gives the desired information in the following sense.
	\begin{itemize}
		\item[$\blacktriangleright$]	If $\mathbf{ p }$ is nonzero, this means that the dinner was paid by one of the cryptographers. We also find out that how much the dinner cost, because $\mathbf{ p }$ is the binary representation of the cost in \texteuro. The identity of the cryptographer who paid can't be inferred from $\mathbf{ p }$, she remains unknown and untraceable.
		\item[$\blacktriangleright$]	If $\mathbf{ p }$ is the zero bit vector $\mathbf{ 0 }$, this means that the dinner was paid by their employer, and not by one of the cryptographers.
	\end{itemize}
	\item	The SQDCP protocol will work even if all the players are in different geographical locations. This is because even if the quantum input registers are spatially separated, they still constitute one composite distributed quantum system due to the strong correlations among their qubits originating from the $\ket{ GHZ_{ n } }$ entanglement. The only difference in the distributed case is that each cryptographer must communicate the obtained measurements to each other cryptographer using pairwise authenticated classical channels.
\end{enumerate}

The whole setup is shown in Figure \ref{fig: The Quantum Circuit for the SQDCP Protocol}. For consistency, all quantum circuits in this paper follow the Qiskit \cite{Qiskit2024} convention in the ordering of their qubits, by placing the least significant qubit at the top of the figure and the most significant at the bottom.

\begin{tcolorbox}
	[
		grow to left by = 1.50 cm,
		grow to right by = 1.50 cm,
		colback = white,	
		enhanced jigsaw,						
		sharp corners,
		toprule = 1.0 pt,
		bottomrule = 1.0 pt,
		leftrule = 0.1 pt,
		rightrule = 0.1 pt,
		sharp corners,
		center title,
		fonttitle = \bfseries
	]
	\begin{figure}[H]
		\centering
		\begin{tikzpicture}[ scale = 1.00 ]
			\begin{yquant}
				nobit AUX_0_0;
				[ name = AGENT_0 ] qubits { $IR_{ 0 }$ } IR_AGENT_0;
				qubit { $OR_{ 0 }$: \ $\ket{ - }$ } OR_AGENT_0;
				nobit AUX_0_1;
				[ name = Dots_0, register/minimum height = 8 mm ] nobit Dots_0;
				[ name = space_0, register/minimum height = 8 mm ] nobit space_0;
				[ name = Dots_n_2, register/minimum height = 8 mm ] nobit Dots_n_2;
				nobit AUX_n_2_0;
				[ name = AGENT_n_2 ] qubits { $IR_{ n - 2 }$ } IR_AGENT_n_2;
				qubit { $OR_{ n - 2 }$: \ $\ket{ - }$ } OR_AGENT_n_2;
				nobit AUX_n_2_1;
				[ name = space_n_2, register/minimum height = 8 mm ] nobit space_n_2;
				nobit AUX_A_0;
				[ name = Alice ] qubits { $AIR$ } AIR;
				qubit { $AOR$: \ $\ket{ - }$ } AOR;
				nobit AUX_A_1;
				nobit AUX_A_2;
				[ name = Ph0, WordBlueDarker, line width = 0.50 mm, label = { [ label distance = 0.20 cm ] north: Initial State } ]
				barrier ( - ) ;
				[ draw = RedPurple!50, fill = RedPurple!50, radius = 0.7 cm ] box { \large \sf{U}$_{ f_{ 0 } }$} (IR_AGENT_0 - OR_AGENT_0);
				[ draw = RedPurple!50, fill = RedPurple!50, radius = 0.7 cm ] box { \large \sf{U}$_{ f_{  n - 2 } }$} (IR_AGENT_n_2 - OR_AGENT_n_2);
				[ draw = RedPurple!50, fill = RedPurple!50, radius = 0.7 cm ] box { \large \sf{U}$_{ f_{ A } }$} (AIR - AOR);
				[ name = Ph1, WordBlueDarker, line width = 0.50 mm, label = { [ label distance = 0.20 cm ] north: Phase 1 } ]
				barrier ( - ) ;
				[ draw = GreenLighter2!50, fill = GreenLighter2!50, radius = 0.5 cm ] box {\large \sf{H}$^{ \otimes m }$} IR_AGENT_0;
				[ draw = GreenLighter2!50, fill = GreenLighter2!50, radius = 0.5 cm ] box {\large \sf{H}$^{ \otimes m }$} IR_AGENT_n_2;
				[ draw = GreenLighter2!50, fill = GreenLighter2!50, radius = 0.5 cm ] box {\large \sf{H}$^{ \otimes m }$} AIR;
				[ name = Ph2, WordBlueDarker, line width = 0.50 mm, label = { [ label distance = 0.20 cm ] north: Phase 2 } ]
				barrier ( - ) ;
				[ draw = white, fill = WordBlueVeryLight, radius = 0.5 cm ] measure IR_AGENT_0;
				[ draw = white, fill = WordBlueVeryLight, radius = 0.5 cm ] measure IR_AGENT_n_2;
				[ draw = white, fill = WordBlueVeryLight, radius = 0.5 cm ] measure AIR;
				[ name = Ph3, WordBlueDarker, line width = 0.50 mm, label = { [ label distance = 0.20 cm ] north: Measurement } ]
				barrier ( - ) ;
				output { $\ket{ \mathbf{ y }_{ 0 } }$ } IR_AGENT_0;
				output { $\ket{ \mathbf{ y }_{ n - 2 } }$ } IR_AGENT_n_2;
				output { $\ket{ \mathbf{ a } }$ } AIR;
				\node [ below = 5.750 cm ] at (Ph0) { $\ket{ \psi_{ 0 } }$ };
				\node [ below = 5.750 cm ] at (Ph1) { $\ket{ \psi_{ 1 } }$ };
				\node [ below = 5.750 cm ] at (Ph2) { $\ket{ \psi_{ 2 } }$ };
				\node [ below = 5.750 cm ] at (Ph3) { $\ket{ \psi_{ f } }$ };
				\node
					[
						charlie,
						scale = 1.50,
						anchor = east,
						left = 0.85 cm of AGENT_0,
						label = { [ label distance = 0.00 cm ] west: $C_0$ }
					]
					(Charlie) { };
				\node
					[
						bob,
						scale = 1.50,
						anchor = center,
						left = 0.50 cm of AGENT_n_2,
						label = { [ label distance = 0.00 cm ] west: $C_{n - 2}$ }
					]
					(Bob) { };
				\node
					[
						alice,
						scale = 1.50,
						anchor = center,
						left = 0.75 cm of Alice,
						label = { [ label distance = 0.00 cm ] west: Alice }
					]
					() { };
				\begin{scope} [ on background layer ]
					\node [ above right = - 0.30 cm and 1.00 cm of space_0, rectangle, fill = WordTurquoiseLighter80!75, text width = 9.00 cm, align = center, minimum height = 10 mm ] { \bf Possibly Spatially Separated };
					\node [ above = 1.75 cm of Bob ] { \LARGE \vdots };
					\node [ above right = - 0.30 cm and 1.00 cm of space_n_2, rectangle, fill = WordTurquoiseLighter80!75, text width = 9.00 cm, align = center, minimum height = 10 mm ] { \bf Possibly Spatially Separated };
				\end{scope}
			\end{yquant}
			\scoped [ on background layer ]
				\draw
				[ MagentaLighter, -, >=stealth, line width = 1.0 mm, decoration = coil, decorate ]
				( $ (Alice.north east) + ( 5 mm, 0 ) $ ) -- ( $ (AGENT_n_2.south east) + ( 5 mm, 0 ) $ );
			\scoped [ on background layer ]
				\draw
				[ MagentaLighter, -, >=stealth, line width = 1.0 mm, decoration = coil, decorate ]
				( $ (AGENT_n_2.north east) + ( 5 mm, 0 ) $ ) -- ( $ (AGENT_0.south east) + ( 5 mm, 0 ) $ );
		\end{tikzpicture}
		\caption{The above figure shows the composite quantum circuit used by the dining cryptographers, exhibiting the individual local circuit Alice and each of her colleagues possess. These local circuits be may spatially separated because they are linked due to entanglement and constitute one composite system. The state vectors $\ket{ \psi_{ 0 } }$, $\ket{ \psi_{ 1 } }$, $\ket{ \psi_{ 2 } }$, and $\ket{ \psi_{ f } }$ describe the evolution of this composite system.}
		\label{fig: The Quantum Circuit for the SQDCP Protocol}
	\end{figure}
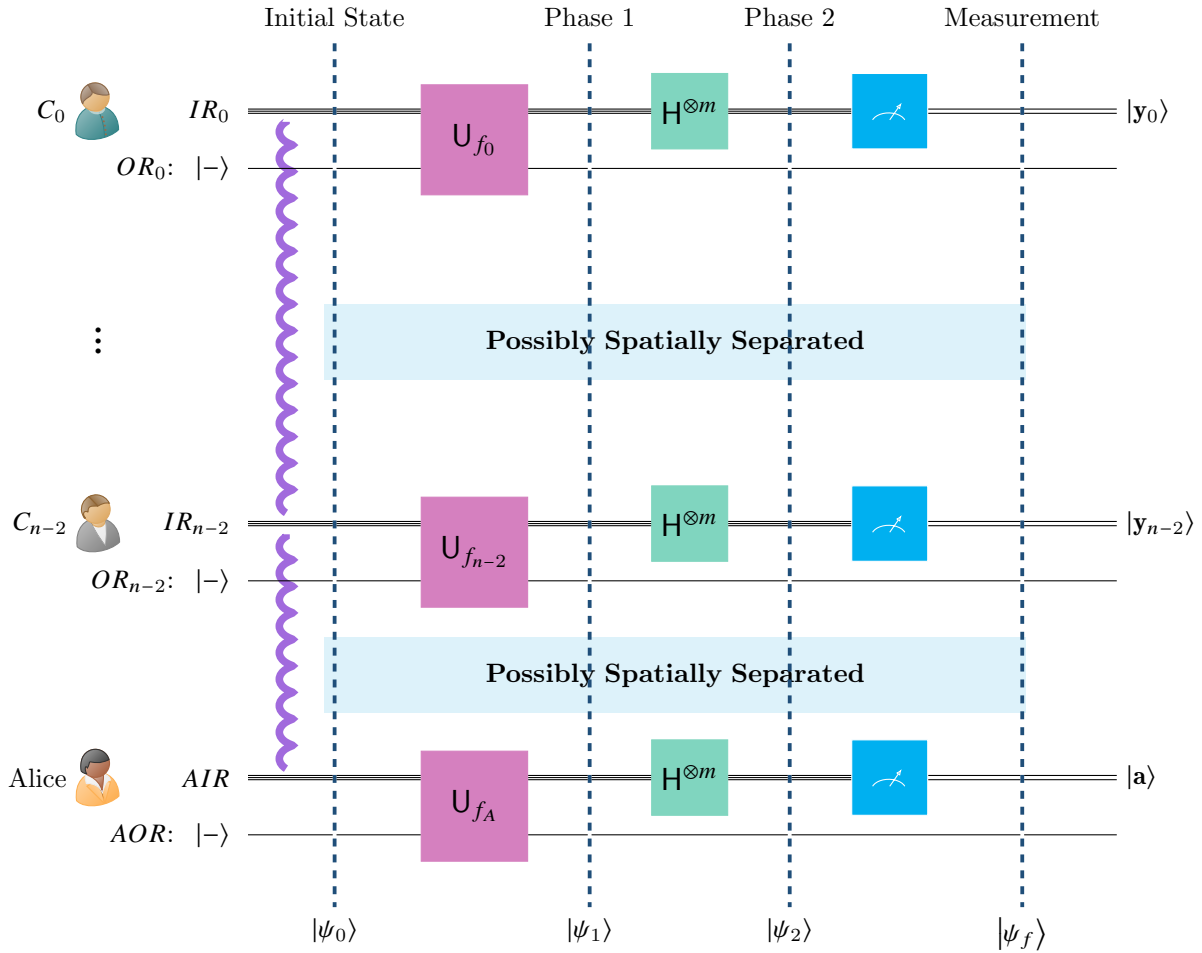
\end{tcolorbox}

In the graphical layout of the above quantum circuit, the notation employed is explained below.

\begin{itemize}
	\item	$AIR$ is Alice's input register.
	\item	$IR_{ i }$ is the input register of cryptographer $C_{ i }$, $0 \leq i \leq n - 2$.
	\item	In total there are $n$ input registers, each containing $m$ qubits. The corresponding qubits in each of the $n$ registers are entangled in the $\ket{ GHZ_{ n } }$ state.
	\item	$AOR$ is Alice's output register.
	\item	$OR_{ i }$ is the output register of cryptographer $C_{ i }$, $0 \leq i \leq n - 2$.
	\item	All output registers contain just a single qubit in the $\ket{ - }$ state.
	\item	$U_{ f_{ A } }$ is Alice's unitary transform.
	\item	$U_{ f_{ i } }$ is the unitary transform of cryptographer $C_{ i }$, $0 \leq i \leq n - 2$.
	\item	$H^{ \otimes m }$ is the $m$-fold Hadamard transform.
\end{itemize}

The initial state of the distributed quantum circuit of Figure \ref{fig: The Quantum Circuit for the SQDCP Protocol} is denoted by $\ket{ \psi_{ 0 } }$, which, using \eqref{eq: m-Fold Extended General GHZ_n State}, can be written as

\begin{align} \label{eq: SQDCP Protocol Initial State}
	\ket{ \psi_{ 0 } }
	=
	\frac { 1 } { \sqrt{ 2^{ m } } }
	\sum_{ \mathbf{ x } \in \mathbb{ B }^{ m } }
	\ket{ - }_{ A }
	\ket{ \mathbf{ x } }_{ A }
	\
	\ket{ - }_{ n - 2 }
	\ket{ \mathbf{ x } }_{ n - 2 }
	\dots
	\ket{ - }_{ 0 }
	\ket{ \mathbf{ x } }_{ 0 }
	\ .
\end{align}

To eliminate any potential ambiguity, we rely on the subscripts $A$, and $i, \ 0 \leq i \leq n - 2$, to indicate whether the kets refer to Alice or cryptographer $i$.

Alice and the other cryptographers start the execution of the SQDCP protocol by acting on their individual quantum circuits via their private unitary transforms $U_{ f_{ A } }$ and $U_{ f_{ i } }$, $0 \leq i \leq n - 2$. By doing so, each one of them can embed the required private information into the entangled state of the composite system. We stress this important fact: any one of them, by using her individual, and local in the distributed scenario, quantum circuit, will encode this private information that must be communicated to the other players into the entangled input registers of the composite, and potentially distributed, circuit.

The unitary transforms $U_{ f_{ A } }$ and $U_{ f_{ i } }$, $0 \leq i \leq n - 2$, are based on private and secret functions $f_{ A }$ and $f_{ i }$, $0 \leq i \leq n - 2$, which are known only to the corresponding cryptographer. Their formal definition is given below.

\begin{align} \label{eq: Alice's Function f_A}
	f_{ A } ( \mathbf{ x } )
	&=
	\mathbf{ p }_{ A } \bullet \mathbf{ x }
	\ , \text{ where }
	\mathbf{ p }_{ A }
	=
	\left\{
	\
	\begin{matrix*}[l]
		\text{the money paid} & \text{if Alice paid for the dinner},
		\\
		\mathbf{ 0 } & \text{if Alice didn't pay for the dinner}.
	\end{matrix*}
	\right.
\end{align}

\begin{align} \label{eq: Cryptographer's Function f_i}
	f_{ i } ( \mathbf{ x } )
	&=
	\mathbf{ p }_{ i } \bullet \mathbf{ x }
	\ , \text{ where }
	\mathbf{ p }_{ i }
	=
	\left\{
	\
	\begin{matrix*}[l]
		\text{the money paid} & \text{if $C_{ i }$ paid for the dinner},
		\\
		\mathbf{ 0 } & \text{if $C_{ i }$ didn't pay for the dinner}.
	\end{matrix*}
	\right.
\end{align}

The unitary transforms $U_{ f }$ follow the typical scheme $U_{ f } \colon \ket{ y } \ket{ \mathbf{ x } }$ $\rightarrow$ $\ket{ y \oplus f ( \mathbf{ x } ) } \ket{ \mathbf{ x } }$. Therefore, in view of \eqref{eq: Alice's Function f_A} and \eqref{eq: Cryptographer's Function f_i}, they can be explicitly written as

\begin{align}
	U_{ f_{ A } }
	&\colon
	\ket{ - }_{ A }
	\ket{ \mathbf{ x } }_{ A }
	\rightarrow
	( - 1 )^{ \mathbf{ p }_{ A } \bullet \mathbf{ x } }
	\
	\ket{ - }_{ A }
	\ket{ \mathbf{ x } }_{ A }
	\ , \text{and}
	\label{eq: Explicit Alice's Unitary Transform U_f_A}
	\\
	U_{ f_{ i } }
	&\colon
	\ket{ - }_{ i }
	\ket{ \mathbf{ x } }_{ i }
	\rightarrow
	( - 1 )^{ \mathbf{ p }_{ i } \bullet \mathbf{ x } }
	\
	\ket{ - }_{ i }
	\ket{ \mathbf{ x } }_{ i }
	\ , \ 0 \leq i \leq n - 2
	\ .
	\label{eq: Explicit Cryptographer's Unitary Transform U_f_i}
\end{align}

The combined effect of the unitary transforms results in the system getting into the state $\ket{ \psi_{ 1 } }$ described below.

\begin{align} \label{eq: SQDCP Protocol Phase 1}
	\ket{ \psi_{ 1 } }
	&=
	\frac { 1 } { \sqrt{ 2^{ m } } }
	\sum_{ \mathbf{ x } \in \mathbb{ B }^{ m } }
	\left(
	U_{ f_{ A } }
	\ket{ - }_{ A }
	\ket{ \mathbf{ x } }_{ A }
	\right)
	\
	\left(
	U_{ f_{ n - 2 } }
	\ket{ - }_{ n - 2 }
	\ket{ \mathbf{ x } }_{ n - 2 }
	\right)
	\
	\dots
	\
	\left(
	U_{ f_{ 0 } }
	\ket{ - }_{ 0 }
	\ket{ \mathbf{ x } }_{ 0 }
	\right)
	\nonumber \\
	&\overset { \eqref{eq: Explicit Alice's Unitary Transform U_f_A}, \eqref{eq: Explicit Cryptographer's Unitary Transform U_f_i} } { = }
	\frac { 1 } { \sqrt{ 2^{ m } } }
	\sum_{ \mathbf{ x } \in \mathbb{ B }^{ m } }
	( - 1 )^{ \mathbf{ p }_{ A } \bullet \mathbf{ x } }
	\ket{ - }_{ A }
	\ket{ \mathbf{x} }_{ A }
	\
	( - 1 )^{ \mathbf{ p }_{ n - 2 } \bullet \mathbf{ x } }
	\ket{ - }_{ n - 2 }
	\ket{ \mathbf{ x } }_{ n - 2 }
	\
	\dots
	\
	( - 1 )^{ \mathbf{ p }_{ 0 } \bullet \mathbf{ x } }
	\ket{ - }_{ 0 }
	\ket{ \mathbf{ x } }_{ 0 }
	\nonumber \\
	&=
	\frac { 1 } { \sqrt{ 2^{ m } } }
	\sum_{ \mathbf{ x } \in \mathbb{ B }^{ m } }
	( - 1 )^{ ( \mathbf{ p }_{ A } \oplus \mathbf{ p }_{ n - 2 } \oplus \dots \oplus \mathbf{ p }_{ 0 } ) \bullet \mathbf{ x } }
	\ket{ - }_{ A }
	\ket{ \mathbf{x} }_{ A }
	\
	\ket{ - }_{ n - 2 }
	\ket{ \mathbf{ x } }_{ n - 2 }
	\
	\dots
	\
	\ket{ - }_{ 0 }
	\ket{ \mathbf{ x } }_{ 0 }
	\ .
\end{align}

Therefore, at the end of Phase 1, the cryptographer who paid for the dinner has embedded the private information known only to her in an untraceable way into the entangled state $\ket{ \psi_{ 1 } }$ of the composite quantum circuit. Now, it remains to decipher this information, so that it becomes known to all other cryptographers. This is explained in the following subsection.

\subsection{Deciphering phase} \label{subsec: Deciphering Phase}

To decipher the embedded private information, Alice and the rest of the cryptographers apply the $m$-fold Hadamard transform to their input registers, as shown in Figure \ref{fig: The Quantum Circuit for the SQDCP Protocol}. Consequently, at the end of Phase 2, the state of the system has become $\ket{ \psi_{ 2 } }$:

\begin{align} \label{eq: SQDCP Protocol Phase 2 - 1}
	\ket{ \psi_{ 2 } }
	=
	\frac { 1 } { \sqrt{ 2^{ m } } }
	\sum_{ \mathbf{ x } \in \mathbb{ B }^{ m } }
	( - 1 )^{ ( \mathbf{ p }_{ A } \oplus \mathbf{ p }_{ n - 2 } \oplus \dots \oplus \mathbf{ p }_{ 0 } ) \bullet \mathbf{ x } }
	\
	\ket{ - }_{ A }
	H^{ \otimes m } \ket{ \mathbf{ x } }_{ A }
	\
	\ket{ - }_{ n - 2 }
	H^{ \otimes m } \ket{ \mathbf{ x } }_{ n - 2 }
	\
	\dots
	\
	\ket{ - }_{ 0 }
	H^{ \otimes m } \ket{ \mathbf{ x } }_{ 0 }
\end{align}

By invoking relation \eqref{eq: Hadamard m-Fold Ket x}, we may analyze $H^{ \otimes m } \ket{ \mathbf{ x } }_{ A }$, $H^{ \otimes m } \ket{ \mathbf{ x } }_{ n - 2 }$, \dots, $H^{ \otimes m } \ket{ \mathbf{ x } }_{ 0 }$ further.

\begin{align*} 
	H^{ \otimes m } \ket{ \mathbf{ x } }_{ A }
	&=
	\frac { 1 } { \sqrt{ 2^{ m } } }
	\sum_{ \mathbf{ a } \in \mathbb{ B }^{ m } }
	( - 1 )^{ \mathbf{ a } \bullet \mathbf{ x } }
	\ket{ \mathbf{ a } }_{ A }
	\\
	H^{ \otimes m } \ket{ \mathbf{ x } }_{ n - 2 }
	&=
	\frac { 1 } { \sqrt{ 2^{ m } } }
	\sum_{ \mathbf{ c }_{ n - 2 } \in \mathbb{ B }^{ m } }
	( - 1 )^{ \mathbf{ c }_{ n - 2 } \bullet \mathbf{ x } }
	\ket{ \mathbf{ c }_{ n - 2 } }_{ n - 2 }
	\\
	&\dots
	\\
	H^{ \otimes m } \ket{ \mathbf{ x } }_{ 0 }
	&=
	\frac { 1 } { \sqrt{ 2^{ m } } }
	\sum_{ \mathbf{ c }_{ 0 } \in \mathbb{ B }^{ m } }
	( - 1 )^{ \mathbf{ c }_{ 0 } \bullet \mathbf{ x } }
	\ket{ \mathbf{ c }_{ 0 } }_{ 0 }
\end{align*}

If we make the above substitutions, $\ket{ \psi_{ 2 } }$ can be cast in an alternative form as shown below.

\begin{align} \label{eq: SQDCP Protocol Phase 2 - 2}
	\ket{ \psi_{ 2 } }
	=
	\frac { 1 } { ( \sqrt{ 2^{ m } )^{ n + 1 } } }
	\sum_{ \mathbf{ a } \in \mathbb{ B }^{ m } }
	\sum_{ \mathbf{ c }_{ n - 2 } \in \mathbb{ B }^{ m } }
	\dots
	\sum_{ \mathbf{ c }_{ 0 } \in \mathbb{ B }^{ m } }
	\sum_{ \mathbf{ x } \in \mathbb{ B }^{ m } }
	( - 1 )^{ ( \mathbf{ p }_{ A } \oplus \mathbf{ p }_{ n - 2 } \oplus \dots \oplus \mathbf{ p }_{ 0 } \oplus \mathbf{ a } \oplus \mathbf{ c }_{ n - 2 } \oplus \dots \oplus \mathbf{ c }_{ 0 } )
		\bullet \mathbf{ x } }
	\nonumber \\
	\ket{ - }_{ A }
	\ket{ \mathbf{ a } }_{ A }
	\
	\ket{ - }_{ n - 2 }
	\ket{ \mathbf{ c }_{ n - 2 } }_{ n - 2 }
	\
	\dots
	\
	\ket{ - }_{ 0 }
	\ket{ \mathbf{ c }_{ 0 } }_{ 0 }
\end{align}

Although the above equation seems complicated, it can be greatly simplified if we apply the characteristic inner product property for zero \eqref{eq: Inner Product Modulo $2$ Property For Zero} and nonzero \eqref{eq: Inner Product Modulo $2$ Property For NonZero} bit vectors. Let us recall what the characteristic inner product property implies in this situation.

\begin{itemize}
	\item[$\blacktriangleright$]	If $\mathbf{ p }_{ A } \oplus \mathbf{ p }_{ n - 2 } \oplus \dots \oplus \mathbf{ p }_{ 0 } \oplus \mathbf{ a } \oplus \mathbf{ c }_{ n - 2 } \oplus \dots \oplus \mathbf{ c }_{ 0 } \neq \mathbf{ 0 }$, or, equivalently, $\mathbf{ a } \oplus \mathbf{ c }_{ n - 2 } \oplus \dots \oplus \mathbf{ c }_{ 0 }$ $\neq$ $\mathbf{ p }_{ A } \oplus \mathbf{ p }_{ n - 2 } \oplus \dots \oplus \mathbf{ p }_{ 0 }$, the sum $\sum_{ \mathbf{ x } \in \mathbb{ B }^{ m } }$ $( - 1 )^{ ( \mathbf{ p }_{ A } \oplus \mathbf{ p }_{ n - 2 } \oplus \dots \oplus \mathbf{ p }_{ 0 } \oplus \mathbf{ a } \oplus \mathbf{ c }_{ n - 2 } \oplus \dots \oplus \mathbf{ c }_{ 0 } ) \bullet \mathbf{ x } }$ $\ket{ - }_{ A }$ $\ket{ \mathbf{ a } }_{ A }$ $\ket{ - }_{ n - 2 } \ket{ \mathbf{ c }_{ n - 2 } }_{ n - 2 }$ $\dots$ $\ket{ - }_{ 0 } \ket{ \mathbf{ c }_{ 0 } }_{ 0 }$ appearing in \eqref{eq: SQDCP Protocol Phase 2 - 2} becomes just $0$.
	\item[$\blacktriangleright$]	If, on the other hand, $\mathbf{ p }_{ A } \oplus \mathbf{ p }_{ n - 2 } \oplus \dots \oplus \mathbf{ p }_{ 0 } \oplus \mathbf{ a } \oplus \mathbf{ c }_{ n - 2 } \oplus \dots \oplus \mathbf{ c }_{ 0 } = \mathbf{ 0 }$, or, equivalently, $\mathbf{ a } \oplus \mathbf{ c }_{ n - 2 } \oplus \dots \oplus \mathbf{ c }_{ 0 }$ $=$ $\mathbf{ p }_{ A } \oplus \mathbf{ p }_{ n - 2 } \oplus \dots \oplus \mathbf{ p }_{ 0 }$, the sum $\sum_{ \mathbf{ x } \in \mathbb{ B }^{ m } }$ $( - 1 )^{ ( \mathbf{ p }_{ A } \oplus \mathbf{ p }_{ n - 2 } \oplus \dots \oplus \mathbf{ p }_{ 0 } \oplus \mathbf{ a } \oplus \mathbf{ c }_{ n - 2 } \oplus \dots \oplus \mathbf{ c }_{ 0 } ) \bullet \mathbf{ x } }$ $\ket{ - }_{ A }$ $\ket{ \mathbf{ a } }_{ A }$ $\ket{ - }_{ n - 2 } \ket{ \mathbf{ c }_{ n - 2 } }_{ n - 2 }$ $\dots$ $\ket{ - }_{ 0 } \ket{ \mathbf{ c }_{ 0 } }_{ 0 }$ becomes $2^{ m }$ $\ket{ - }_{ A }$ $\ket{ \mathbf{ a } }_{ A }$ $\ket{ - }_{ n - 2 } \ket{ \mathbf{ c }_{ n - 2 } }_{ n - 2 }$ $\dots$ $\ket{ - }_{ 0 } \ket{ \mathbf{ c }_{ 0 } }_{ 0 }$.
\end{itemize}

The above explanation allows us to cast $\ket{ \psi_{ 2 } }$ in the following reduced form.

\begin{align} \label{eq: SQDCP Protocol Phase 2 - 3}
	\ket{ \psi_{ 2 } }
	=
	\frac { 1 } { ( \sqrt{ 2^{ m } )^{ n - 1 } } }
	\sum_{ \mathbf{ a } \in \mathbb{ B }^{ m } }
	\sum_{ \mathbf{ c }_{ n - 2 } \in \mathbb{ B }^{ m } }
	\dots
	\sum_{ \mathbf{ c }_{ 0 } \in \mathbb{ B }^{ m } }
	\ket{ - }_{ A }
	\ket{ \mathbf{ a } }_{A}
	\
	\ket{ - }_{ n - 2 }
	\ket{ \mathbf{ c }_{ n - 2 } }_{ n - 2 }
	\
	\dots
	\
	\ket{ - }_{ 0 }
	\ket{ \mathbf{ c }_{ 0 } }_{0}
	\ ,
\end{align}

where

\begin{align} \label{eq: Fundamental Correlation Property}
	\mathbf{ a }
	\oplus
	\mathbf{ c }_{ n - 2 }
	\oplus
	\dots
	\oplus
	\mathbf{ c }_{ 0 }
	=
	\mathbf{ p }_{ A }
	\oplus
	\mathbf{ p }_{ n - 2 }
	\oplus
	\dots
	\oplus \mathbf{ p }_{ 0 }
	\ .
\end{align}

Using the terminology introduced in \cite{Ampatzis2023} and \cite{Andronikos2023}, we refer to the relation \eqref{eq: Fundamental Correlation Property} as the Fundamental Correlation Property that intertwines the input registers of the dining cryptographers. This relation is the aftermath of the initial entanglement among all the input registers. At the end of Phase 2, the cryptographer who paid for the dinner has embedded the relevant private information in the global state of the composite quantum circuit, which has caused this constraint on the contents of the input registers.

The quantum part of the SQDCP protocol is over when the cryptographers measure their input registers in the computational basis. By this action, the state of the composite system collapses to the final state $\ket{ \psi_{ f } }$ that has the following form.

\begin{align}
	\ket{ \psi_{ f } }
	&=
	\ket{ - }_{ A }
	\ket{ \mathbf{ a } }_{A}
	\
	\ket{ - }_{ n - 2 }
	\ket{ \mathbf{ c }_{ n - 2 } }_{ n - 2 }
	\
	\dots
	\
	\ket{ - }_{ 0 }
	\ket{ \mathbf{ c }_{ 0 } }_{0}
	\ ,
	\text{ where }
	\label{eq: SQDCP Protocol Final Measurement}
	\\
	&\mathbf{ a }
	\oplus
	\mathbf{ c }_{ n - 2 }
	\oplus
	\dots
	\oplus
	\mathbf{ c }_{ 0 }
	=
	\mathbf{ p }_{ A }
	\oplus
	\mathbf{ p }_{ n - 2 }
	\oplus
	\dots
	\oplus \mathbf{ p }_{ 0 }
	\label{eq: SQDCP Protocol Final Sum}
\end{align}

Let us emphasize that all the above equations hold true in both the localized and the distributed versions of the SQDCP protocol because their validity stems from the initial entanglement among all the input registers. As long as entanglement is present, the distance among the dining cryptographers plays no role.

The protocol completes the task of actually decrypting the embedded private information through the following steps.

\begin{enumerate} [ left = 0.50 cm, labelsep = 1.00 cm ]
	\renewcommand\labelenumi{(\textbf{D}$_\theenumi$)}
	\item	Every cryptographer communicates to every other cryptographer the measured contents of her input register. That is, Alice sends $\mathbf{ a }$ to her $n - 1$ cryptographer colleagues, and each $C_{ i }$, $0 \leq i \leq n - 2$, sends $\mathbf{ c }_{ i }$ to Alice and every other cryptographer.
	\item	In a localized setting, this step is quite trivial. In a distributed setting, it is also easily achievable, as it only requires the use of pairwise authenticated classical communication channels.
	\item	At this point, every player knows all bit vectors $\mathbf{ a }, \mathbf{ c }_{ n - 2 }, \dots, \mathbf{ c }_{ 0 }$. This allows each cryptographer to compute the modulo $2$ sum $\mathbf{ a } \oplus \mathbf{ c }_{ n - 2 } \oplus \dots \oplus \mathbf{ c }_{ 0 }$, which, according to \eqref{eq: SQDCP Protocol Final Sum}, produces the modulo $2$ sum $\mathbf{ p }_{ A } \oplus \mathbf{ p }_{ n - 2 } \oplus \dots \oplus \mathbf{ p }_{ 0 }$.
	\item	The modulo $2$ sum $\mathbf{ p }_{ A } \oplus \mathbf{ p }_{ n - 2 } \oplus \dots \oplus \mathbf{ p }_{ 0 }$ conveys the information the cryptographers wanted to uncover in the first place. Here's why.
	\begin{itemize}
		\item	In case none of the cryptographers paid for the dinner, then, according to (\textbf{PIE}$_{ 2 }$) and (\textbf{PIE}$_{ 3 }$), $\mathbf{ p }_{ A } = \mathbf{ p }_{ n - 2 } = \dots = \mathbf{ p }_{ 0 } = \mathbf{ 0 }$. Consequently, their modulo $2$ sum is $\mathbf{ 0 }$, which means that the computed modulo $2$ sum $\mathbf{ a }, \mathbf{ c }_{ n - 2 }, \dots, \mathbf{ c }_{ 0 }$ is also $\mathbf{ 0 }$. Hence, the cryptographers infer that the dinner was paid by their employer.
		\item	In case $C_{ i }$, $0 \leq i \leq n - 2$, paid for the dinner a certain amount of money, then, considering (\textbf{PIE}$_{ 1 }$) -- (\textbf{PIE}$_{ 3 }$), $\mathbf{ p }_{ i }$, the binary representation of this amount, is nonzero, whereas $\mathbf{ p }_{ A }$ and all other $\mathbf{ p }_{ j }$, $j \neq i$, are zero. In turn, their modulo $2$ sum is $\mathbf{ 0 }$, which means that the computed modulo $2$ sum $\mathbf{ a }, \mathbf{ c }_{ n - 2 }, \dots, \mathbf{ c }_{ 0 }$ is $\mathbf{ p }_{ i }$. Therefore, the cryptographers infer that it was one of them who paid for the dinner, and, as an added bonus, they also get to know how much the dinner cost. Obviously, the same argument goes verbatim in case it was Alice who paid for the dinner.
	\end{itemize}
	\item	The above explanation also shows that the original source of the information remains unknown and untraceable. The private information, be it $\mathbf{ p }_{ A }$ or some $\mathbf{ p }_{ i }$, $0 \leq i \leq n - 2$, has been absorbed into the sum $\mathbf{ p }_{ A } \oplus \mathbf{ p }_{ n - 2 } \oplus \dots \oplus \mathbf{ p }_{ 0 }$ and there is no way that it can be retrieved.
\end{enumerate}

\begin{example} \label{xmp: Alice, Bob, Charlie & Dave Use the SQDCP Protocol} \
	The present example is a continuation of our previous Example \ref{xmp: Alice, Bob, Charlie & Dave's Dinner}. It doesn't matter whether they are physically together around the same table, or if they are in different geographical locations dining virtually, the SQDCP Protocol will go through in both settings. We also recall that their private quantum input registers contain $4$ qubits, each. The corresponding qubits in the $4$ quantum registers form a quadruple entangled in the $\ket{ GHZ_{ 4 } }$, according to the entanglement distribution scheme outlined in Definition \ref{def: Entanglement Distribution Scheme}.

	First, let us consider the case where one of the cryptographers, Alice, paid for the dinner. If Alice paid say $12$ \texteuro, then she embeds the binary representation of $12$, namely, $\mathbf{ p }^{ A } = 1100$ in the entangled state of the composite circuit. This is easily achieved via CNOT gates. If we implement the general quantum circuit shown in Figure \ref{fig: The Quantum Circuit for the SQDCP Protocol} in Qiskit, we end up with the specific implementation depicted in Figure \ref{fig: SQDCP Example Alice Paid Quantum Circuit}. By measuring their input registers, Alice, Bob, Charlie, and Dave get one of the $2^{ 16 } = 65536$ equiprobable outcomes.

\begin{tcolorbox}
	[
		grow to left by = 1.50 cm,
		grow to right by = 1.50 cm,
		colback = white,			
		enhanced jigsaw,			
		sharp corners,
		boxrule = 0.01 pt,
		toprule = 0.01 pt,
		bottomrule = 0.01 pt
	]
	\begin{figure}[H]
		\centering
		\includegraphics[ scale = 0.44, angle = 90, trim = {0 0 0cm 0}, clip ]{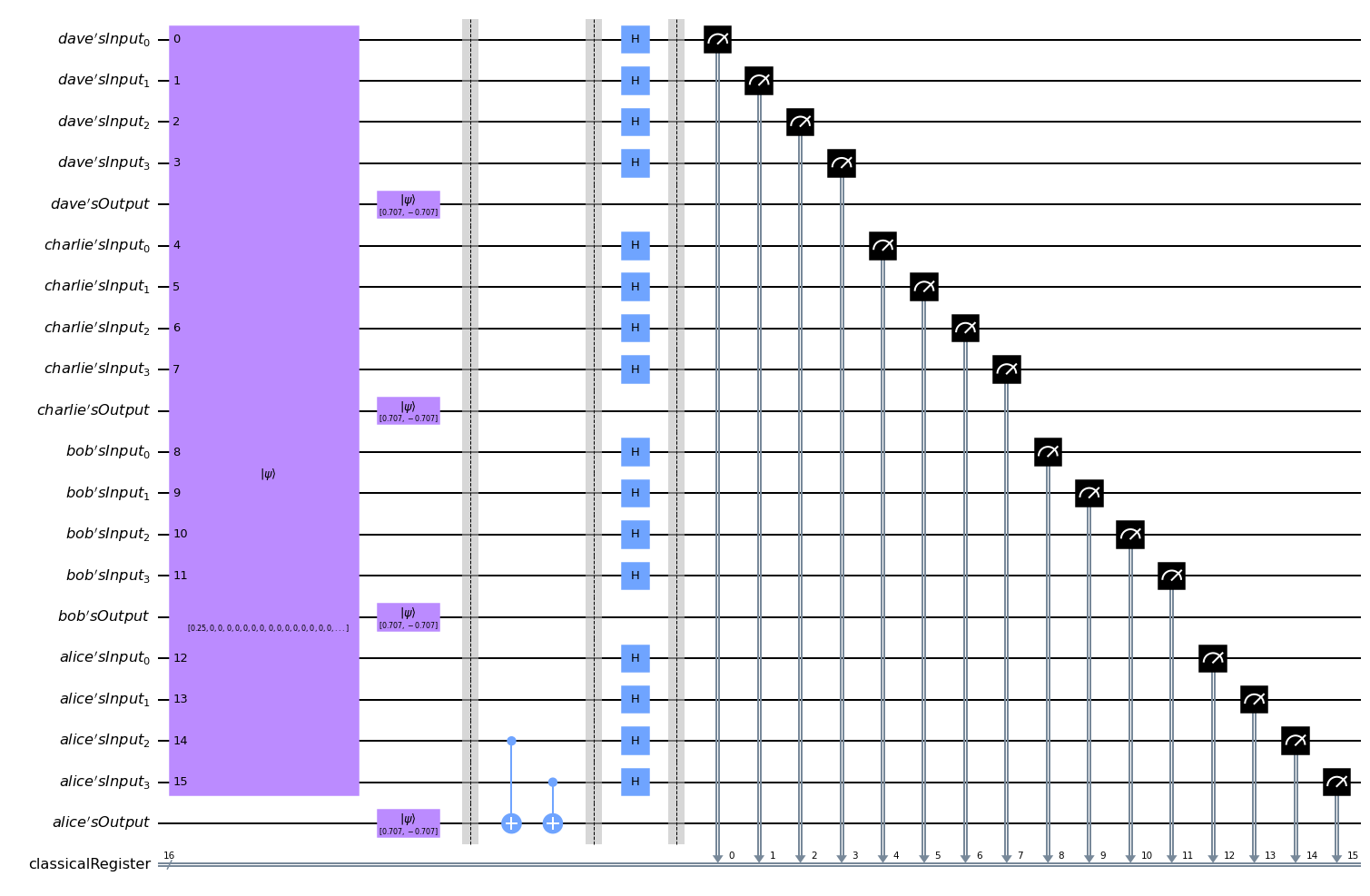}
		\caption{The above quantum circuit simulates the SQDCP protocol corresponding to the case where Alice paid for the dinner, as outlined in Example \ref{xmp: Alice, Bob, Charlie & Dave Use the SQDCP Protocol}.}
		\label{fig: SQDCP Example Alice Paid Quantum Circuit}
	\end{figure}
\end{tcolorbox}

\begin{tcolorbox}
	[
		grow to left by = 0.00 cm,
		grow to right by = 0.00 cm,
		colback = white,			
		enhanced jigsaw,			
		sharp corners,
		boxrule = 0.01 pt,
		toprule = 0.01 pt,
		bottomrule = 0.01 pt
	]
	\begin{figure}[H]
		\centering
		\includegraphics[ scale = 0.35, angle = 90, trim = {0cm 0cm 0cm 0cm}, clip ]{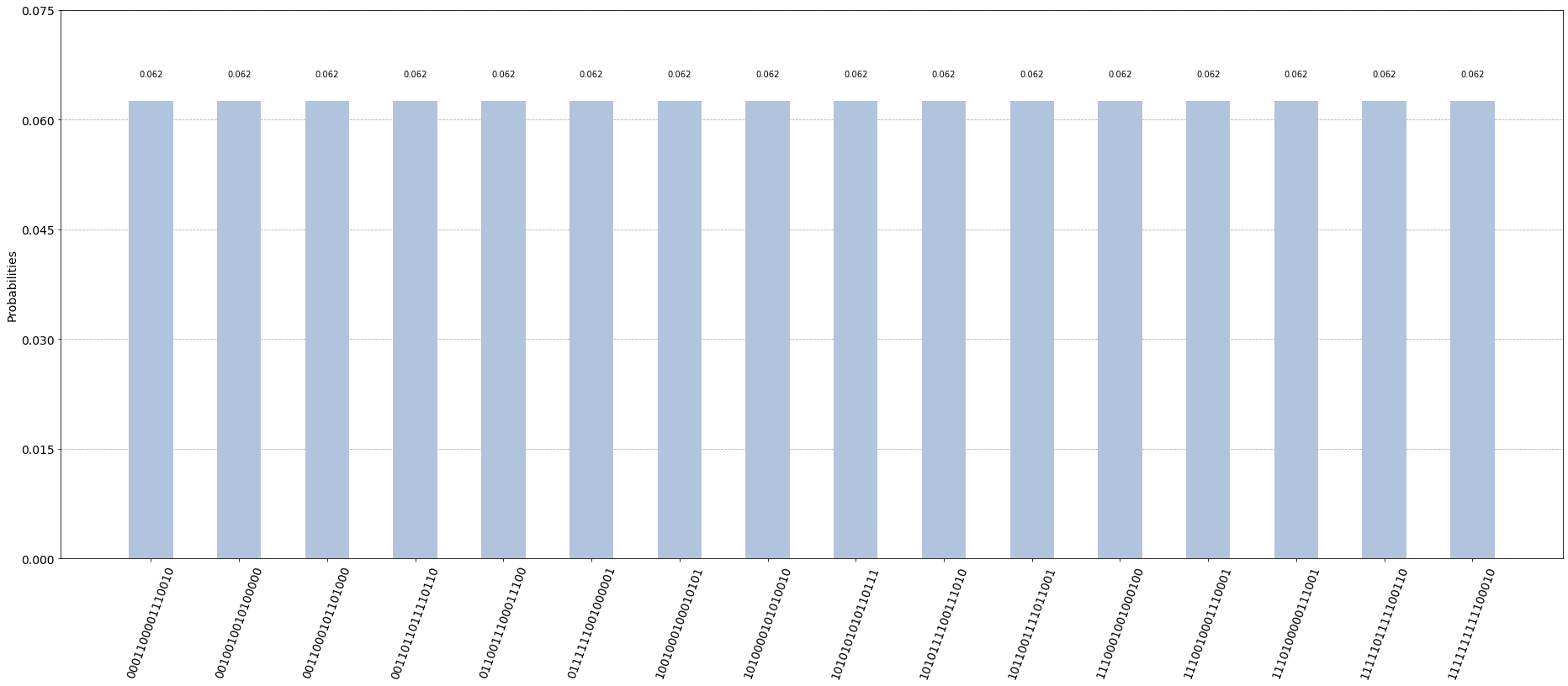}
		\caption{Some of the possible measurements and their corresponding probabilities for the circuit of Figure \ref{fig: SQDCP Example Alice Paid Quantum Circuit}.}
		\label{fig: SQDCP Example Alice Paid Measurement Outcomes}
	\end{figure}
\end{tcolorbox}

\begin{tcolorbox}
	[
		grow to left by = 1.50 cm,
		grow to right by = 1.50 cm,
		colback = white,			
		enhanced jigsaw,			
		sharp corners,
		boxrule = 0.01 pt,
		toprule = 0.01 pt,
		bottomrule = 0.01 pt
	]
	\begin{figure}[H]
		\centering
		\includegraphics[ scale = 0.44, angle = 90, trim = {0 0 0cm 0}, clip ]{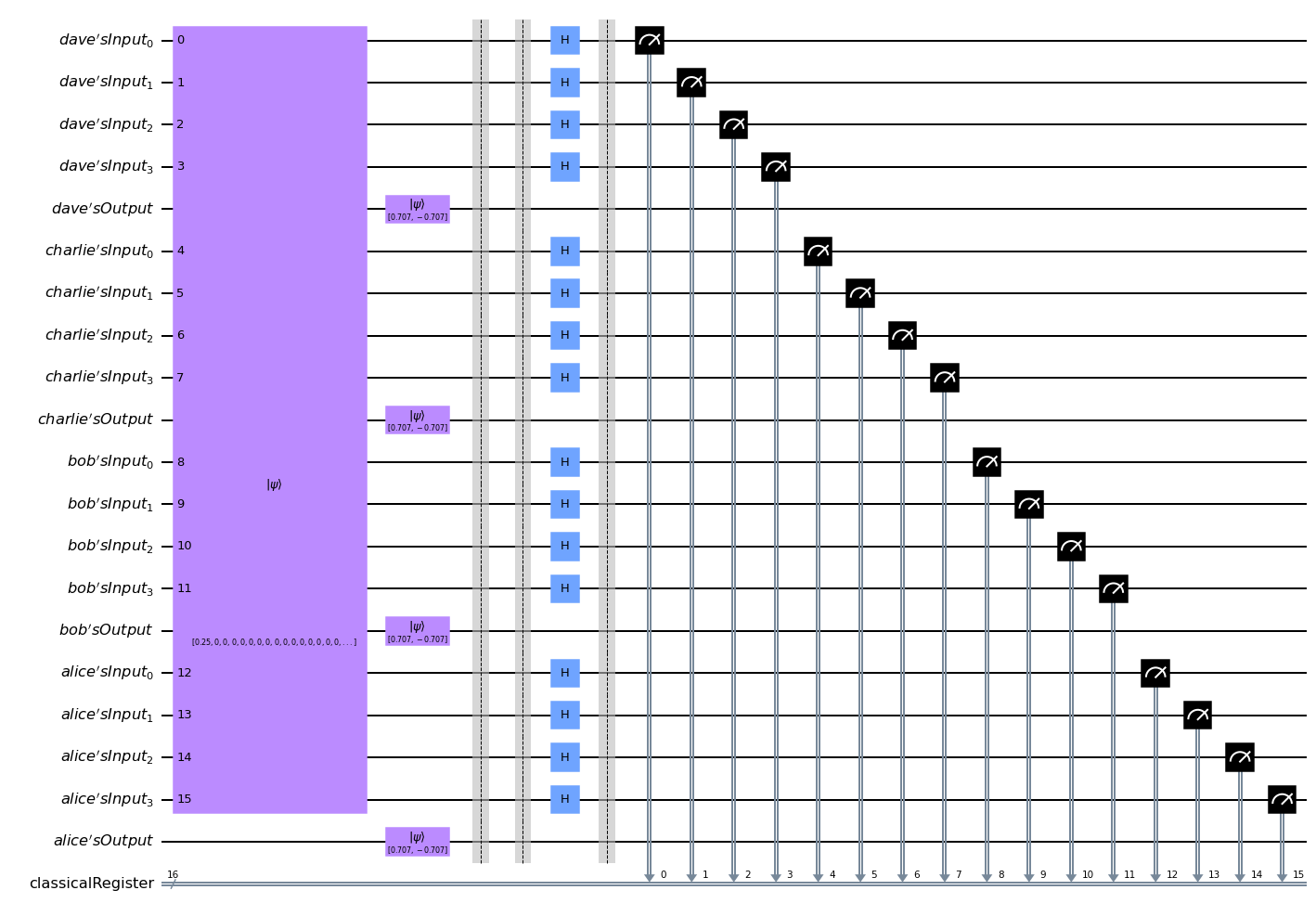}
		\caption{The above quantum circuit simulates the SQDCP protocol corresponding to the case where none of the four cryptographers paid for the dinner, as outlined in Example \ref{xmp: Alice, Bob, Charlie & Dave Use the SQDCP Protocol}.}
		\label{fig: SQDCP Example None Paid Quantum Circuit}
	\end{figure}
\end{tcolorbox}

\begin{tcolorbox}
	[
		grow to left by = 0.00 cm,
		grow to right by = 0.00 cm,
		colback = white,			
		enhanced jigsaw,			
		sharp corners,
		boxrule = 0.01 pt,
		toprule = 0.01 pt,
		bottomrule = 0.01 pt
	]
	\begin{figure}[H]
		\centering
		\includegraphics[ scale = 0.35, angle = 90, trim = {0cm 0cm 0cm 0cm}, clip ]{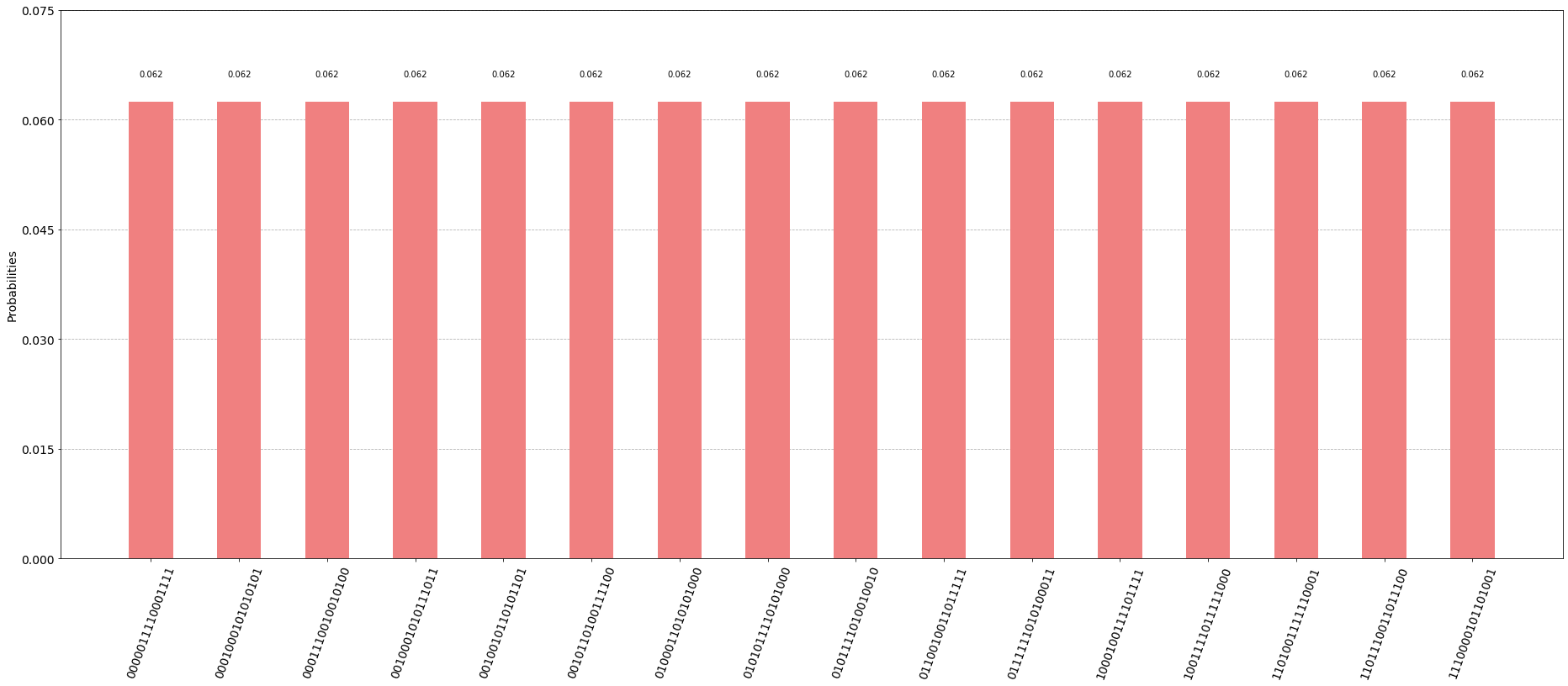}
		\caption{Some of the possible outcomes and their corresponding probabilities for the circuit of Figure \ref{fig: SQDCP Example None Paid Quantum Circuit}.}
		\label{fig: SQDCP Example None Paid Measurement Outcomes}
	\end{figure}
\end{tcolorbox}

	For obvious technical limitations, we can't show all these outcomes, since this would result in an unintelligible figure. Hence, we depict only 16 of them in Figure \ref{fig: SQDCP Example Alice Paid Measurement Outcomes}. It is straightforward to check that every possible outcome satisfies the Fundamental Correlation Property and verifies equations \eqref{eq: Fundamental Correlation Property} and \eqref{eq: SQDCP Protocol Final Sum}. For example, we may examine the label of the first bar of the histogram contained in Figure \ref{fig: SQDCP Example Alice Paid Measurement Outcomes}, which is $0001 \ 1000 \ 0111 \ 0010$. This means that upon measurement the contents of Alice, Bob, Charlie, and Dave's input register are $\mathbf{ a } = 0001$, $\mathbf{ b } = 1000$, $\mathbf{ c } = 0111$, and $\mathbf{ d } = 0010$, respectively. These contents are shared among the four cryptographers, according to (\textbf{D}$_{ 1 }$) and (\textbf{D}$_{ 2 }$), and become common knowledge to all of them. Finally, they XOR them together to uncover the secret information, i.e., $\mathbf{ p } = \mathbf{ a } \oplus \mathbf{ b } \oplus \mathbf{ c } \oplus \mathbf{ d } = 1100$, which leads them to infer that one of them paid $12$ \texteuro for the dinner. The crucial thing is that neither the measured contents $\mathbf{ a }$, $\mathbf{ b }$, $\mathbf{ c }$, and $\mathbf{ d }$ of the input registers, nor the final private information $\mathbf{ p }$ can reveal the identity of the cryptographer who paid the bill.

	Let us also briefly examine the case where none of the cryptographers paid for the dinner. In such a situation all cryptographers embed the zero bit vector in the entangled state of the composite circuit, i.e., $\mathbf{ p }_{ A } = \mathbf{ p }_{ B } = \mathbf{ p }_{ C } = \mathbf{ p }_{ D } = 0000$. The quantum circuit in this case is shown in Figure \ref{fig: SQDCP Example None Paid Quantum Circuit}. Again, it is very easy to ascertain that every possible outcome satisfies the Fundamental Correlation Property and verifies equations \eqref{eq: Fundamental Correlation Property} and \eqref{eq: SQDCP Protocol Final Sum}. For example, we may examine the label of the third bar of the histogram contained in Figure \ref{fig: SQDCP Example None Paid Measurement Outcomes}, which is $0001 \ 1100 \ 1001 \ 0100$. This means that the contents of Alice, Bob, Charlie, and Dave's input register are $\mathbf{ a } = 0001$, $\mathbf{ b } = 1100$, $\mathbf{ c } = 1001$, and $\mathbf{ d } = 0100$, respectively. These contents are shared among the four cryptographers, according to (\textbf{D}$_{ 1 }$) and (\textbf{D}$_{ 2 }$), and become common knowledge to all of them. Finally, they XOR them together to uncover the secret information, i.e., $\mathbf{ p } = \mathbf{ a } \oplus \mathbf{ b } \oplus \mathbf{ c } \oplus \mathbf{ d } = 0000$, from which they deduce that none of them paid for the dinner, so it must have been their employer. \hfill $\triangleleft$
\end{example}

\section{Discussion and conclusions} \label{sec: Discussion and Conclusions}

This work introduces a novel entanglement-based protocol for solving the Dining Cryptographers problem. The proposed protocol relies on maximally entangled $\ket{ GHZ_{ n } }$ tuples to achieve its goal. The main incentive was to offer scalability in terms of both the number of cryptographers $n$ and the amount of anonymous information it conveys, measured by the number $m$ of qubits contained in each quantum register. The number of cryptographers $n$ can be any arbitrary large positive integer. In addition to scalability in the number of participants, our protocol can seamlessly scale in terms of the amount of anonymous information it transmits. Originally, the Dining Cryptographers Problem involved only one bit of information, namely whether one of the cryptographers paid for the dinner. In the proposed SQDCP protocol, the number $m$ of qubits in each register can also be any arbitrarily large positive integer. This number reflects the amount of information that can be transmitted. For instance, the cryptographer who actually paid the bill may also disclose how much it cost or when the arrangement was made, etc.

Another noteworthy feature of the protocol introduced in this work is its ability to address both a localized and distributed version of the Dining Cryptographers Problem. The localized scenario involves all cryptographers being together at the same spatial location, i.e., at the same restaurant, at a specific point in time. The distributed scenario involves cryptographers being in different spatial locations, but virtually having dinner at the same time.

Finally, in terms of the actual implementation of the protocol, we point out that all cryptographers employ private quantum circuits that are identical. This achieves a completely modular quantum system, with all distinct modules being the same. Furthermore, each private quantum circuit utilizes only the ubiquitous Hadamard and CNOT quantum gates, making them easily implemented on contemporary quantum computers.

\bibliographystyle{ieeetr}
\bibliography{SQDCP}

\end{document}